\documentclass{article}
\usepackage[utf8]{inputenc}

\RequirePackage[OT1]{fontenc}
\RequirePackage{amsthm,amsmath}
\RequirePackage{natbib,enumerate}
\RequirePackage[citecolor=blue,urlcolor=blue]{hyperref}
\usepackage[dvipsnames,svgnames]{xcolor}
\usepackage{bbold}
\usepackage[normalem]{ulem}
\usepackage{tikz}
\usepackage{comment}
\usepackage{pgfplots}
\usepackage{filecontents}
\usepackage{booktabs}
\usepackage{hyperref}
\usepackage[T1]{fontenc}
\usepackage[ruled,vlined]{algorithm2e}
\usepackage{thmtools} 
\usepackage{subcaption}
\usepackage{float}
\usepackage{subcaption}
\usepackage{caption}
\usepackage{enumitem}
\usepackage{thmtools, thm-restate}
 
\usepackage{tikz}
\usepackage{soul}
\usetikzlibrary{shapes,arrows,positioning}
\usepackage[format=plain,labelfont=bf,textfont=it, font=small]{caption}

\usepackage{graphicx}
\usepackage{amsfonts}
\usepackage{wrapfig}
\usepackage{comment}
\usepackage{pifont}
\usepackage{tabularx}

\newcommand{\R}{\ensuremath{{\mathbb R}}}
\newcommand{\E}{\ensuremath{{\mathbb E}}}

\newcommand{\G}{\ensuremath{{\mathcal G}}}

\usepackage{lscape}

\usepackage{dsfont}
\usepackage{caption}
\usepackage{float}

\newcommand{\distas}[1]{\mathbin{\overset{#1}{\kern\z\sim}}}%
\newsavebox{\mybox}\newsavebox{\mysim}
\newcommand{\distras}[1]{%
  \savebox{\mybox}{\hbox{\kern3pt$\scriptstyle#1$\kern3pt}}%
  \savebox{\mysim}{\hbox{$\sim$}}%
  \mathbin{\overset{#1}{\kern\z@\resizebox{\wd\mybox}{\ht\mysim}{$\sim$}}}%
}

\newcommand{\nrofmethods}{73 }
\newcommand{\nrofmethodsnum}{70 }

\newcommand{\nrofmethodscat}{23 }
\newcommand{\nrofmethodboth}{21 }
\newcommand{\nrofdatasets}{24 }

\usepackage{varioref}
\usepackage{cleveref}
\usepackage{longtable}

\defcitealias{allergens_datachallenge}{SFA and ABN (2024)}

\usepackage[toc]{appendix}

\usepackage[textwidth=3cm]{todonotes}

\numberwithin{equation}{section}
\theoremstyle{plain}

\title{Do we Need Dozens of Methods for Real World Missing Value
Imputation?}
\author{}
\date{\today}
\author{Krystyna Grzesiak$ ^1$, Christophe Muller$^2$, Julie Josse$^3$, Jeffrey Näf$^4$ \\
$ ^1$Faculty of Mathematics and Computer Science, University of Wroc\l{}aw\\
$ ^2$Department of Statistics, University of Oxford\\
$ ^3$Inria, PreMeDICaL Team, University of Montpellier\\
$ ^4$Research Institute for Statistics and Information Science\\
University of Geneva\\
}

\begin{document}

\maketitle

\begin{abstract}
    Missing values pose a persistent challenge in modern data science. Consequently, there is an ever-growing number of publications introducing new imputation methods in various fields. While many studies compare imputation approaches, they often focus on a limited subset of algorithms and evaluate performance primarily through pointwise metrics such as RMSE, which are not suitable to measure the preservation of the true data distribution. In this work, we provide a systematic benchmarking method based on the idea of treating imputation as a distributional prediction task. We consider a large number of algorithms and, for the first time, evaluate them not only on synthetic missing mechanisms, but also on real-world missingness scenarios, using the concept of Imputation Scores. Finally, while the focus of previous benchmark has often been on numerical data, we also consider mixed data sets in our study. The analysis overwhelmingly confirms the superiority of iterative imputation algorithms, especially the methods implemented in the \texttt{mice} \textsf{R} package.
\end{abstract}

\paragraph{Keywords:} imputation, missing at random, distributional prediction, proper scores 


\section{Introduction}

Missing values represent a pervasive challenge that significantly affects research methodologies and scientific investigations across virtually all disciplines. Imputation of missing values has a long tradition and is often the preferred approach to deal with this issue, as it yields one or more complete data sets that can subsequently be used for analysis.

Consequently, a wealth of imputation methods have been developed in many different fields. These methods are typically categorized into two main approaches: joint modeling, which imputes the data using a single (implicit or explicit) model, and fully conditional specification (FCS), which involves training a separate model for each variable \citep{FCS_Van_Buuren2007, VANBUUREN2018}. The most famous instance of the latter is the multiple imputation by chained equations (mice) approach \citep{mice}. Examples of joint modeling include using parametric distributions \citep{schafer1997analysis}, and more recently, Generative Adversarial Network (GAN)-based \citep{GAIN, directcompetitor1, directcompetitor2} and Variational Autoencoder (VAE)-based methods \citep{MIWAE,VAE1,VAE2, VAE3}. 
This methodological expansion, driven by advances in machine learning, has significantly increased the number of available imputation algorithms. As a result, practitioners aiming to impute a given dataset face numerous options. While there exists a substantial literature comparing imputation methods, existing benchmarks typically consider a limited subset of available algorithms, and, as we will argue, their recommendations may be misleading. 

In this work, our objective is to implement a large-scale benchmarking of imputation methods using the principles developed in \citet{näf2025good}. In particular, we refrain from using standard pointwise metrics such as Root Mean Squared Error (RMSE) or Mean Absolute Error (MAE)—predominantly used in earlier benchmark studies—as these metrics evaluate point predictions (e.g., mean or median imputation) and are thus limited in assessing the true data distribution. As already noted in \citet[Chapter 2.6]{VANBUUREN2018}, \textit{“Imputation is not prediction”} and metrics such as RMSE and MAE can be highly misleading indicators of imputation success, potentially contributing to an unjustified emphasis on single imputation methods such as \texttt{missForest} \citep{stekhoven2012missforest}. Instead, we assess how well the distribution of the imputed data aligns with that of the original data. That is, for artificially introduced missing values we compare the distributions of full and imputed data using the energy distance \citep{EnergyDistance}, following the approach proposed by \citet{näf2025good}. This allows for a more holistic and principled evaluation of imputation performance. To this end, we provide an extensive analysis of \nrofmethods imputation algorithms, using \nrofdatasets data sets of different sizes and 2 artificially created missing value mechanisms, Missing Completely at Random (MCAR) and Missing at Random (MAR), with varying percentages of missing values. These two mechanisms are standard, using the \textit{ampute} method of \citet{mice}, which allows control over the missingness pattern and permits a direct comparison between imputed and original data. 

While artifically creating missing values in fully available data sets is the main approach in benchmarking studies, the data are rarely missing in such a neat way. Thus, we also evaluate the imputation methods on real missingness; that is, on data sets with naturally occurring missing values. To the best of our knowledge, this is the first general purpose imputation benchmark to include such data sets. Since the true values are unknown, we cannot rely on the energy distance or traditional metrics such as RMSE to assess the imputation performance. Instead, we rely on the Imputation Scores (I-Scores) introduced by \citet{ImputationScores} and improved in \citet{näf2025Iscore}, which provides a proper scoring rule under a certain MAR assumption. The I-Score framework offers a way to compare imputation quality without access to the ground truth, making it uniquely suited for real-world missingness. Although it comes with the caveat of relying on a MAR assumption to reliably identify the best method, it opens up new possibilities with mechanisms that are introduced by actual data collection, as opposed to the often simplistic artificial mechanisms commonly used in the literature.

The title of the paper was inspired by \citet{delgadoclassifier} in the context of classification. As in their paper, we will reduce the many imputation methods employed in practice to a few that tend to work well over a wide range of settings. However, we note that imputation is a complex problem, and depending on the data, the missingness mechanism, and the goal of the analysis, some imputation methods might be preferable to those recommended here. A very simple example could be the imputation of a high-dimensional data set for which joint methods like GAIN of \citet{GAIN} and MIWAE of \citet{MIWAE}, which perform comparatively poorly in this analysis, might be preferred for computational reasons. Another example is prediction with missing values in the features. In this case it has been shown that, at least asymptotically, simple mean or zero imputation may suffice \citep{Marine_2021_Prediction, Josse2024consistency, lemorvan2025imputation}.

A key goal of our work is the adherence to the FAIR principles proposed in \citet{fair}, ensuring that all aspects of the benchmark are Findable, Accessible, Interoperable, and Reusable. To this end, we provide a fully reproducible and extensible benchmarking framework. This includes open access to all data sets, code, configuration files, and results, enabling researchers to both replicate our findings and easily incorporate new methods or data sets into the benchmarking system. All code used to generate the results in this study, along with configuration files, evaluation scripts, and preprocessed data sets, is publicly available in a dedicated GitHub repository: \url{github.com/ChristopheMuller/benchmark}. We also shared selected datasets and methods from this study on the \texttt{R-miss-tastic} platform \citep{mayer2024rmisstasticunifiedplatformmissing}.

The paper proceeds as follows: This section concludes with comparison to earlier benchmarks and an example motivating our approach in Section \ref{Sec_Introexample}. In Section \ref{Sec_Setup} we then introduce our methodology in detail, including considered methods, handling of categorical variables and the way we measure the success of imputation methods under real and artificial missing values. Section \ref{Sec_Discussion} then summarizes our findings and discusses our final recommendations as well as the limitations of the study. Finally, Sections \ref{Sec_ArtificialMissing} and \ref{Sec_Realmissing} present the results in detail for artificial and real missingness, respectively.

\subsection{Related Work and Contributions} \label{Sec_Context}


Previous benchmarking was carried out mainly using (versions of) RMSE and MAE and was often limited in the number of compared methods. As a consequence, many benchmarking papers recommend methods that simply impute by (conditional) expectation and do not try to recover the distribution of the true data. Moreover, despite its limitations,  RMSE remains widely used in the introduction of new imputation methods \citep[see, e.g.][]{stekhoven2012missforest, GAIN, egert2021dima}, often painting a misleading picture. Consider, for instance, the well-cited paper by \citet{GAIN}. The authors acknowledge the importance of recovering the true underlying data distribution and provide elegant results supporting their method's ability to do so. However, they then demonstrate that their method outperforms previous ones in terms of RMSE. Given the known properties of RMSE, this paradoxically suggests that their method may actually recover the data distribution \emph{less accurately} than competing approaches.

\begin{table}[ht]
\centering
\begingroup\scriptsize
\begin{tabular}{l|cc|ccc}
  \toprule
\textbf{Benchmark} & \multicolumn{2}{c|}{\textbf{Real Missingness}} & \multicolumn{3}{c}{\textbf{Artificial Missingness}} \\
 & Distributional & Downstream & Pointwise & Distributional & Downstream \\
\midrule
This Work & \checkmark & -- & -- & \checkmark & -- \\ 
  \citet{benchmark_shadbahr2023impact} & -- & \checkmark & \checkmark & \checkmark & \checkmark \\ 
  \citet{benchmark_jager2021} & -- & -- & \checkmark & -- & \checkmark \\ 
  \citet{benchmark_woznica2020does} & -- & \checkmark & -- & -- & -- \\ 
  \citet{benchmark_knn_adv1} & -- & -- & \checkmark & -- & -- \\ 
  \citet{benchmark_metabolomics1} & -- & -- & \checkmark & -- & -- \\ 
  \citet{benchmark_kyureghian2011missing} & -- & -- & \checkmark & -- & \checkmark \\ 
  \citet{benchmark_joel2024performance} & -- & -- & \checkmark & -- & \checkmark \\ 
  \citet{benchmark_pereira2024imputation} & -- & -- & \checkmark & -- & -- \\ 
  \citet{benchmark_alam2023investigation} & -- & -- & \checkmark & -- & \checkmark \\ 
  \citet{benchmark_sun2023deep} & -- & -- & \checkmark & -- & -- \\ 
  \citet{benchmark_pavelchek2023imputation} & -- & -- & \checkmark & -- & -- \\ 
  \citet{benchmark_ge2023simulation} & -- & -- & \checkmark & -- & -- \\ 
  \citet{benchmark_seu2022intelligent} & -- & -- & \checkmark & -- & -- \\ 
  \citet{benchmark_deforth2024performance} & -- & -- & -- & -- & \checkmark \\ 
  \citet{benchmark_xu2020} & -- & -- & \checkmark & -- & -- \\ 
  \citet{benchmark_poulos2018missing} & -- & -- & -- & -- & \checkmark \\ 
  \citet{benchmark_getz2023performance} & -- & -- & -- & -- & \checkmark \\ 
  \citet{benchmark_wang2022deep} & -- & -- & \checkmark & -- & -- \\ 
  \citet{benchmark_wongkamthong2023comparative} & -- & -- & -- & -- & \checkmark \\ 
  \citet{benchmark_miao2022experimental} & -- & -- & \checkmark & -- & -- \\ 
  \citet{imputomics} & -- & -- & \checkmark & -- & -- \\ 
  \citet{benchmark_junninen2004methods} & -- & -- & \checkmark & -- & -- \\ 
   \bottomrule
\end{tabular}
\endgroup
\caption{Overview of benchmark studies and evaluation criteria. ``Pointwise'' refers to the use of measures such as RMSE and MAE to compare imputed and complete data sets. ``Downstream'' refers to tasks such as classification on the imputed data sets.}
\label{tab:benchmark_overview}
\end{table}

To the best of our knowledge, the only other benchmarks comparing the joint distribution of imputed and complete data are \citet{hyperimpute} and \citet{benchmark_shadbahr2023impact}. In particular, the latter uses an algorithm based on the sliced-Wasserstein distance \citep{slicedWasserstein}. Although we believe this is a step in the right direction, their benchmark is limited to a small number of data sets, considers only MCAR missingness, and focuses primarily on the performance of imputation methods in downstream classification tasks. In this specific context, accurately replicating the underlying data distribution may be less critical, as suggested by the growing body of literature on prediction with missing data. Specifically,  \citet{Marine_2021_Prediction, Josse2024consistency} show that even trivial imputations of features can still recover the Bayes predictor asymptotically. The results in \citet{benchmark_shadbahr2023impact} thus lend empirical credence to this asymptotic result, although this link was not made in their paper. In contrast, we are more interested in general-purpose imputation, for which capturing the correct distribution is essential. Moreover, compared to their somewhat convoluted way of measuring distances between imputed and real data sets using the sliced Wasserstein distance, we use a simpler energy distance-based approach that is fast to calculate and simple to implement. As such, our analysis is more related to \citet{hyperimpute} who compare methods using the Wasserstein distance between imputed and real data sets. However, since the authors introduce a new imputation framework, the focus seems more on showing the performance of this new methodology than on an extensive benchmarking. Finally, we study more data sets and a wider range of missingness mechanisms than both earlier papers.

\begin{figure}[H]
    \makebox[\linewidth]{\includegraphics[width=1.3\linewidth]{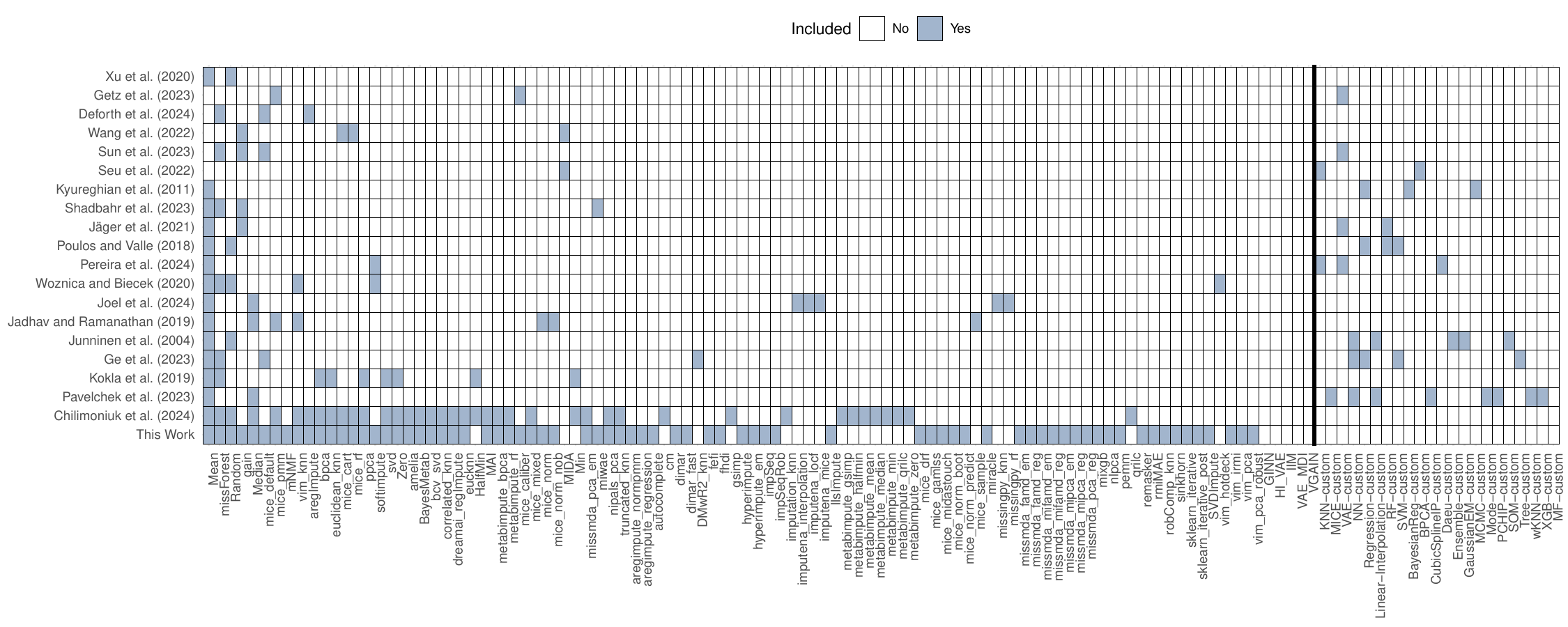}}
    \caption{Methods compared by selected benchmarks from the literature. We note that some benchmarks implement methods themselves instead of relying on publicly available implementations. We refer to these as ``custom''  methods.}
    \label{fig:benchmarks_comparison}
\end{figure}

Similarly, while there have been previous evaluations on downstream (prediction) tasks for real-world missing data (see \Cref{tab:benchmark_overview}), to the best of our knowledge, this is the first attempt at ranking imputation methods on real data sets for general-purpose imputation. While using a downstream task for comparison is valid when that specific task is of interest, it does not reflect the general performance of an imputation method. In fact, relying on an arbitrary set of downstream tasks for general assessment is not only limited in scope, but can also be misleading: For example, as mentioned above, a naive imputation by the mean might lead to good performance if prediction is considered as a downstream task, even though this tends to heavily distort the original distribution. On the other hand, such a distortion of the distribution will have a dramatic effect on measures of distributional distance. \Cref{fig:benchmarks_comparison} provides a comparison of the current paper with previous benchmarks. 


Overall, our analysis allows us to reduce more than 70 imputation methods to a handful of methods that appear to work well in a wide range of settings. We would argue that these methods should become the new baseline for comparison. Our code base also allows other researchers to plug in their own methods to compare with others. This should help streamline and guide the development of new imputation methods.



\subsection{Motivation: Imputation is not Prediction}\label{Sec_Introexample}

\begin{figure}
    \centering
    \includegraphics[width=1 \linewidth]{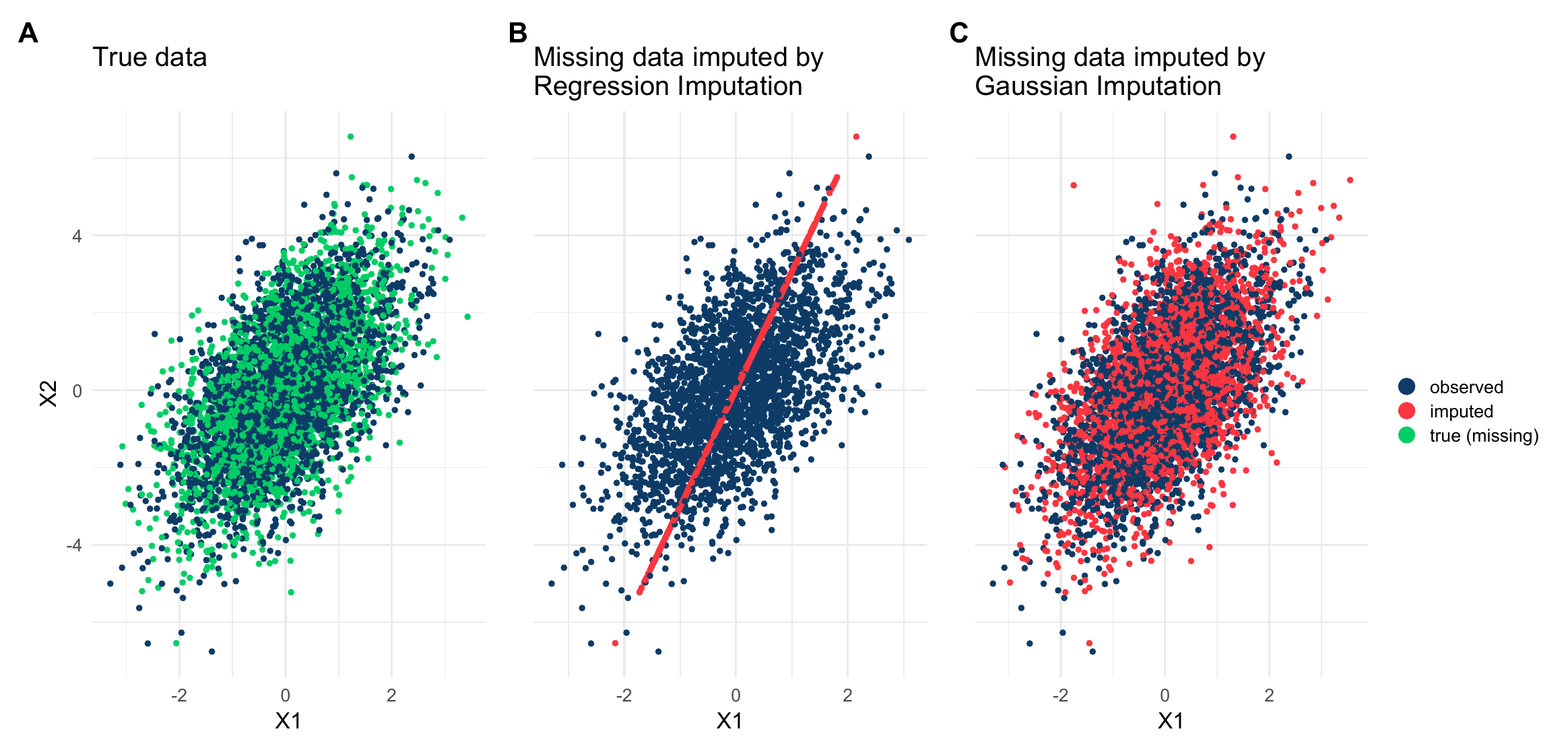}
    \caption{5000 observations of the bivariate Gaussian Example with around 50\% MCAR missing values in $X_1$.(A) Original observations (observed and missing), (B) Imputation by fitting a regression model and imputing the prediction (\texttt{mice\_norm.predict}), (C) Imputation by fitting a regression model and imputing by drawing from conditional Gaussian distribution (\texttt{mice\_norm}).}
    \label{fig:MotivationExample}
\end{figure}

It has long been known, see e.g., \citet[Chapter 2.6]{VANBUUREN2018}, that imputation is not a prediction task in the sense that one should not impute the \emph{most likely} value. However, measures like the RMSE or MAE favor imputations that impute (conditional) mean or median values, such as the regression imputation, missForest from \citet{stekhoven2012missforest} and k-NN imputation \citep{knnreference}. While this can be valid if the goal is predicting a univariate variable $Y$ \citep{Josse2024consistency,Marine_2021_Prediction}, it may bias further analysis. Figure \ref{fig:MotivationExample} illustrates these problems. We simulated 5000 observations of a bivariate Gaussian example and added around 50\% of MCAR missing values to $X_1$. Then we imputed using both the regression imputation and the Gaussian imputation. The former simply imputes the conditional mean based on regression estimates, while the latter also fits a regression, but then imputes by drawing from the conditional Gaussian distribution. Visually, it is already quite clear that the regression imputation can be problematic, as it is not able to capture the geometry of the data. On the other hand, the Gaussian imputation is able to recreate the distribution quite accurately. This discrepancy can affect parameter estimation in downstream analyses. For instance, despite the high sample size, regressing $X_2$ onto the imputed $X_1$ leads to a bias for the regression imputation (estimated value $\approx$ 1.5, true value = 1). In fact, in this specific (MCAR) example, it would be better to just perform the regression on complete cases (estimated value $\approx$ 1.03) than to use regression imputation. On the other hand, the Gaussian imputation is able to retrieve the true value of 1 quite accurately (estimated value $\approx$ 1.01). Despite these disadvantages of the regression imputation, it scores better in terms of RMSE (0.60) than the Gaussian imputation (0.84). 

Thus RMSE can be highly misleading as a measure of imputation quality, with potential detrimental effects on downstream tasks.
Instead, in trying to judge the quality of imputation methods, we might choose a downstream task directly. In this case, regressing $X_2$ onto the imputed $X_1$ would reveal that the Gaussian imputation is superior. However, had we instead chosen to regress $X_1$ onto $X_2$ as a downstream method to test the imputations, it would seem as if the two imputation methods perform roughly the same. Thus the ``wrong'' choice of downstream method might hide the differences in the imputation quality. Although such an approach may be valid if only one particular task is of interest, it is not an ideal way to judge the quality of an imputation method.

Finally, using the energy distance to measure the distance between imputed and true distribution, immediately reveals the Gaussian imputation to be better. Similarly, the energy-I-Score used below recognizes that the Gaussian imputation is better without access to the original complete data set.


\section{Methodology}\label{Sec_Setup}

\paragraph{Considered Methods} 

\Cref{tab:methods} summarizes all methods compared in our benchmark, along with their corresponding implementations. For the sake of convenience, we relied on the wrappers from \texttt{imputomics} \citep{imputomics}, an R package and web application that provides an implementation of over 40 missing value imputation algorithms, \texttt{hyperimpute} \citep{hyperimpute}, a Python package that similarly provides implementations for a range of algorithms, and \texttt{mice} \citep{mice, R_mice} implementing a large number of mice approaches, for some methods. Unless specified otherwise, all methods are employed with their default hyperparameter settings.  Some methods were excluded either because they were designed for specialized imputation tasks (e.g., censored data), because a functional implementation could not be found, or simply due to computational constraints.

\begingroup\scriptsize
\begin{longtable}{rlll}
  \hline
 & Methods & Languages & Implementations \\ 
  \hline
1 & \textbf{amelia} \citep{amelia} & \textsf{R} & \texttt{CRAN:Amelia} \citep{R_Amelia}$^{\ast}$ \\ 
  2 & \textbf{aregImpute} \citep{VANBUUREN2018} & \textsf{R} & \texttt{CRAN:Hmisc} \citep{R_Hmisc} \\ 
  3 & \textbf{aregimpute\_normpmm} \citep{VANBUUREN2018} & \textsf{R} & \texttt{CRAN:Hmisc} \citep{R_Hmisc} \\ 
  4 & \textbf{aregimpute\_regression} \citep{VANBUUREN2018} & \textsf{R} & \texttt{CRAN:Hmisc} \citep{R_Hmisc} \\ 
  5 & \textbf{autocomplete} \citep{an2023deep} & \texttt{Python} & \texttt{\href{https://github.com/sriramlab/AutoComplete}{Git:sriramlab/AutoComplete}} \\ 
  6 & \textbf{BayesMetab} \citep{shah2019bayesmetab} & \textsf{R} & \texttt{Provided by authors}$^{\ast}$ \\ 
  7 & \textbf{bcv\_svd} \citep{knnreference} & \textsf{R} & \texttt{CRAN:bcv} \citep{R_bcv}$^{\ast}$ \\ 
  8 & \textbf{bpca} \citep{oba2003bayesian} & \textsf{R} & \texttt{bioconductor:pcaMethods} \citep{R_pcaMethods}$^{\ast}$ \\ 
  9 & \textbf{correlated\_knn} \citep{wei2018gsimp} & \textsf{R} & \texttt{\href{https://github.com/WandeRum/GSimp}{Git:WandeRum/GSimp}}$^{\ast}$ \\ 
  10 & \textbf{dimar} \citep{egert2021dima} & \textsf{R} & \texttt{\href{https://github.com/kreutz-lab/DIMAR}{Git:kreutz-lab/DIMAR}} \\ 
  11 & \textbf{dimar\_fast} \citep{egert2021dima} & \textsf{R} & \texttt{\href{https://github.com/kreutz-lab/DIMAR}{Git:kreutz-lab/DIMAR}} \\ 
  12 & \textbf{dreamai\_regImpute} \citep{ma2020dreamai} & \textsf{R} & \texttt{\href{https://github.com/WangLab-MSSM/DreamAI}{Git:WangLab-MSSM/DreamAI}}$^{\ast}$ \\ 
  13 & \textbf{eucknn} \citep{knnreference} & \textsf{R} & \texttt{\href{https://github.com/WandeRum/GSimp}{Git:WandeRum/GSimp}}$^{\ast}$ \\ 
  14 & \textbf{euclidean\_knn} \citep{knnreference} & \textsf{R} & \texttt{bioconductor:impute} \citep{R_impute}$^{\ast}$ \\ 
  15 & \textbf{fefi} \citep{im2018fhdi} & \textsf{R} & \texttt{CRAN:FHDI} \citep{R_FHDI} \\ 
  16 & \textbf{fhdi} \citep{im2018fhdi} & \textsf{R} & \texttt{CRAN:FHDI} \citep{R_FHDI} \\ 
  17 & \textbf{gain} \citep{GAIN} & \texttt{Python} & \texttt{\href{https://github.com/jsyoon0823/GAIN}{Git:jsyoon0823/GAIN}}$^{\dagger}$ \\ 
  18 & \textbf{hyperimpute} \citep{hyperimpute} & \texttt{Python} & \texttt{\href{https://github.com/vanderschaarlab/hyperimpute}{Git:vanderschaarlab/hyperimpute}}$^{\dagger}$ \\ 
  19 & \textbf{hyperimpute\_em} \citep{garcia2010pattern} & \texttt{Python} & \texttt{\href{https://github.com/vanderschaarlab/hyperimpute}{Git:vanderschaarlab/hyperimpute}}$^{\dagger}$ \\ 
  20 & \textbf{impSeq} \citep{impSeqMethod} & \textsf{R} & \texttt{CRAN:rrcovNA} \citep{R_rrcovNA} \\ 
  21 & \textbf{impSeqRob} \citep{impSeqMethod} & \textsf{R} & \texttt{CRAN:rrcovNA} \citep{R_rrcovNA} \\ 
  22 & \textbf{llsImpute} \citep{kim2005missing} & \textsf{R} & \texttt{bioconductor:pcaMethods} \citep{R_pcaMethods} \\ 
  23 & \textbf{MAI} \citep{dekermanjian2022mechanism} & \textsf{R} & \texttt{bioconductor:MAI} \citep{R_MAI}$^{\ast}$ \\ 
  24 & \textbf{mean} & \textsf{R} & \texttt{CRAN:base} \citep{R_base}$^{\ast}$ \\ 
  25 & \textbf{median} & \textsf{R} & \texttt{CRAN:base} \citep{R_base}$^{\ast}$ \\ 
  26 & \textbf{metabimpute\_bpca} \citep{stacklies2007pcamethods} & \textsf{R} & \texttt{\href{https://github.com/BeanLabASU/metabimpute}{Git:BeanLabASU/metabimpute}}$^{\ast}$ \\ 
  27 & \textbf{metabimpute\_rf} \citep{Tang2017-on} & \textsf{R} & \texttt{\href{https://github.com/BeanLabASU/metabimpute}{Git:BeanLabASU/metabimpute}} \\ 
  28 & \textbf{mice\_caliber} \citep{VANBUUREN2018} & \textsf{R} & \texttt{CRAN:CALIBERrfimpute} \citep{R_CALIBERrfimpute} \\ 
  29 & \textbf{mice\_cart} \citep{VANBUUREN2018} & \textsf{R} & \texttt{CRAN:mice} \citep{R_mice} \\ 
  30 & \textbf{mice\_default} \citep{VANBUUREN2018} & \textsf{R} & \texttt{CRAN:mice} \citep{R_mice} \\ 
  31 & \textbf{mice\_drf} \citep{näf2025good} & \textsf{R} & \texttt{\href{https://github.com/KrystynaGrzesiak/miceDRF}{Git:KrystynaGrzesiak/miceDRF}} \\ 
  32 & \textbf{mice\_gamlss} \citep{de2016multiple} & \textsf{R} & \texttt{CRAN:ImputeRobust} \citep{R_ImputeRobust} \\ 
  33 & \textbf{mice\_midastouch} \citep{VANBUUREN2018} & \textsf{R} & \texttt{CRAN:mice} \citep{R_mice} \\ 
  34 & \textbf{mice\_mixed} \citep{VANBUUREN2018} & \textsf{R} & \texttt{CRAN:missCompare} \citep{R_missCompare} \\ 
  35 & \textbf{mice\_norm} \citep{VANBUUREN2018} & \textsf{R} & \texttt{CRAN:mice} \citep{R_mice} \\ 
  36 & \textbf{mice\_norm\_boot} \citep{VANBUUREN2018} & \textsf{R} & \texttt{CRAN:mice} \citep{R_mice} \\ 
  37 & \textbf{mice\_norm\_nob} \citep{VANBUUREN2018} & \textsf{R} & \texttt{CRAN:mice} \citep{R_mice} \\ 
  38 & \textbf{mice\_norm\_predict} \citep{VANBUUREN2018} & \textsf{R} & \texttt{CRAN:mice} \citep{R_mice} \\ 
  39 & \textbf{mice\_pmm} \citep{VANBUUREN2018} & \textsf{R} & \texttt{CRAN:mice} \citep{R_mice} \\ 
  40 & \textbf{mice\_rf} \citep{VANBUUREN2018} & \textsf{R} & \texttt{CRAN:mice} \citep{R_mice} \\ 
  41 & \textbf{miracle} \citep{kyono2021miracle} & \texttt{Python} & \texttt{\href{https://github.com/vanderschaarlab/MIRACLE}{Git:vanderschaarlab/MIRACLE}}$^{\dagger}$ \\ 
  42 & \textbf{missForest} \citep{stekhoven2012missforest} & \textsf{R} & \texttt{CRAN:missForest} \citep{R_missForest} \\ 
  43 & \textbf{missmda\_famd\_em} \citep{josse2016missmda} & \textsf{R} & \texttt{CRAN:missMDA} \citep{R_missMDA} \\ 
  44 & \textbf{missmda\_famd\_reg} \citep{josse2016missmda} & \textsf{R} & \texttt{CRAN:missMDA} \citep{R_missMDA} \\ 
  45 & \textbf{missmda\_mifamd\_em} \citep{josse2016missmda} & \textsf{R} & \texttt{CRAN:missMDA} \citep{R_missMDA} \\ 
  46 & \textbf{missmda\_mifamd\_reg} \citep{josse2016missmda} & \textsf{R} & \texttt{CRAN:missMDA} \citep{R_missMDA} \\ 
  47 & \textbf{missmda\_mipca\_em} \citep{josse2016missmda} & \textsf{R} & \texttt{CRAN:missMDA} \citep{R_missMDA} \\ 
  48 & \textbf{missmda\_mipca\_reg} \citep{josse2016missmda} & \textsf{R} & \texttt{CRAN:missMDA} \citep{R_missMDA} \\ 
  49 & \textbf{missmda\_pca\_em} \citep{josse2016missmda} & \textsf{R} & \texttt{CRAN:missMDA} \citep{R_missMDA} \\ 
  50 & \textbf{missmda\_pca\_reg} \citep{josse2016missmda} & \textsf{R} & \texttt{CRAN:missMDA} \citep{R_missMDA} \\ 
  51 & \textbf{miwae} \citep{mattei2018missiwae} & \texttt{Python} & \texttt{\href{https://github.com/pamattei/miwae}{Git:pamattei/miwae}}$^{\dagger}$ \\ 
  52 & \textbf{mixgb} \citep{deng2024multiple} & \textsf{R} & \texttt{CRAN:mixgb} \citep{R_mixgb} \\ 
  53 & \textbf{mNMF} \citep{xu2021nmf} & \textsf{R} & \texttt{\href{https://github.com/freeoliver-jing/NMF}{Git:freeoliver-jing/NMF}}$^{\ast}$ \\ 
  54 & \textbf{nipals\_pca} \citep{stacklies2007pcamethods} & \textsf{R} & \texttt{bioconductor:pcaMethods} \citep{R_pcaMethods}$^{\ast}$ \\ 
  55 & \textbf{nlpca} \citep{stacklies2007pcamethods} & \textsf{R} & \texttt{bioconductor:pcaMethods} \citep{R_pcaMethods} \\ 
  56 & \textbf{pemm} \citep{chen2014penalized} & \textsf{R} & \texttt{CRAN:PEMM} \citep{R_pemm}$^{\ast}$ \\ 
  57 & \textbf{ppca} \citep{stacklies2007pcamethods} & \textsf{R} & \texttt{bioconductor:pcaMethods} \citep{R_pcaMethods}$^{\ast}$ \\ 
  58 & \textbf{random} & \textsf{R} & \texttt{CRAN:base} \citep{R_base} \\ 
  59 & \textbf{sinkhorn} \citep{muzellec2020missing} & \texttt{Python} & \texttt{\href{https://github.com/BorisMuzellec/MissingDataOT}{Git:BorisMuzellec/MissingDataOT}}$^{\dagger}$ \\ 
  60 & \textbf{remasker} \citep{du2023remasker} & \texttt{Python} & \texttt{\href{https://github.com/tydusky/remasker}{Git:tydusky/remasker}} \\ 
  61 & \textbf{rmiMAE} \citep{kumar2019new} & \textsf{R} & \texttt{\href{https://github.com/NishithPaul/missingImputation}{Git:NishithPaul/missingImputation}} \\ 
  62 & \textbf{robComp\_knn} \citep{knnreference} & \textsf{R} & \texttt{CRAN:robCompositions} \citep{R_robcompositions} \\ 
  63 & \textbf{sklearn\_iterative} \citep{VANBUUREN2018} & \texttt{Python} & \texttt{PyPI:scikit-learn} \citep{sklearn_package} \\ 
  64 & \textbf{sklearn\_iterative\_post} \citep{VANBUUREN2018} & \texttt{Python} & \texttt{PyPI:scikit-learn} \citep{sklearn_package} \\ 
  65 & \textbf{softimpute} \citep{hastie2015matrix} & \textsf{R} & \texttt{CRAN:softImpute} \citep{R_softimpute}$^{\ast}$ \\ 
  66 & \textbf{svd} \citep{stacklies2007pcamethods} & \textsf{R} & \texttt{bioconductor:pcaMethods} \citep{R_pcaMethods}$^{\ast}$ \\ 
  67 & \textbf{SVDImpute} \citep{stacklies2007pcamethods} & \textsf{R} & \texttt{CRAN(archive):imputation} \citep{R_imputation} \\ 
  68 & \textbf{truncated\_knn} \citep{wei2018gsimp} & \textsf{R} & \texttt{\href{https://github.com/WandeRum/GSimp}{Git:WandeRum/GSimp}}$^{\ast}$ \\ 
  69 & \textbf{vim\_irmi} \citep{templ2011iterative} & \textsf{R} & \texttt{CRAN:VIM} \citep{R_VIM} \\ 
  70 & \textbf{vim\_knn} \citep{knnreference} & \textsf{R} & \texttt{CRAN:VIM} \citep{R_VIM}$^{\ast}$ \\ 
  71 & \textbf{vim\_pca} \citep{serneels2008principal} & \textsf{R} & \texttt{CRAN:VIM} \citep{R_VIM} \\ 
  72 & \textbf{vim\_pca\_robust} \citep{serneels2008principal} & \textsf{R} & \texttt{CRAN:VIM} \citep{R_VIM} \\ 
  73 & \textbf{zero} & \textsf{R} & \texttt{CRAN:base} \citep{R_base}$^{\ast}$ \\ 
   \hline
\hline
\caption{List of imputation methods included in our work. $^{\ast}$Implemented via imputomics wrapper. $^{\dagger}$Implemented via hyperimpute wrapper. } 
\label{tab:methods}
\end{longtable}
\endgroup


\paragraph{Missing Value Mechanisms} We use the \texttt{produce\_NA} function from the R-miss-tastic platform \citep{mayer2021rmisstastic}\footnote{function available at \url{https://rmisstastic.netlify.app/how-to/generate/misssimul}}, for the case of artificially created missing values. This function is a wrapper around the widely used \texttt{ampute} routine from the \texttt{mice} package \citep{ampute}, and thus provides standard missingness mechanisms commonly used in the literature \citep[see, e.g.,][]{ampute_func_1,ampute_func_2,ampute_func_3}. As these amputations are random procedures, we apply $2$ replicates of each amputation to mitigate this randomness. For categorical variables, the amputation process was constrained to ensure that no category level was entirely removed from any variable. In the second part of our benchmark, we consider data with real missing values, where the missingness mechanism is unknown. 

\paragraph{Missing Proportion} For artificially generated missing values, the missing proportion is defined as the total number of missing entries divided by $n\times p$, where $n$ denotes the number of observations and $p$ the number of variables. We consider missing proportions of 0.1, 0.2, and 0.3, applied uniformly across all matrix elements. For the case of real missing values, the missing proportions of the data sets are given in column four of Table \ref{tab:datasets}.

\paragraph{Categorical Variables} Let $X$ be a categorical variable taking values in a set of discrete values $\G$. For instance, $\G$ could be a set of names of categories, or $\G$ might consist of natural numbers indicating a class, or a set of binary vectors representing a one-hot encoding of a given categorical variable. We loosely define an imputation method to be capable of dealing with categorical variables if it exclusively produces values for $X$ that belong to the set of valid categories ($\G$). For instance, \texttt{mice\_rf} regresses a categorical variable with missing values on all other variables and then draws from the set of categories for imputation. Thus, \texttt{mice\_rf} is capable of dealing with categorical variables. On the other hand, \texttt{mean} imputation for numerically encoded categorical variable may produce values outside $\G$ (e.g. $2.5$ for $\G=\{2,3\}$). Thus, \texttt{mean} imputation is not capable of dealing with categorical variables. 

Crucially, a method's ability to handle categorical variables often depends on the specific data representation used. For instance, while \texttt{mice\_rf} can deal with categorical variables when they are defined as factors in \textsf{R}, it will fail to do so under one-hot encoding of the categories. In this case, the iterative imputation of \texttt{mice\_rf} may produce categories which are not part of $\G$. For instance, the mice iterations may produce the vector $(1,1)$, when $\G=\{(1,0), (0,1)\}$. Conversely, some methods that are not explicitly designed for categorical variables may still work correctly under certain representations. We will thus consider four possible representations: either coding the categories as integer numbers, as factor numbers in \textsf{R}, factor characters in \textsf{R} or as one-hot encoded numerical variables. A total of \nrofmethodscat methods in this benchmark work with at least one representation, as displayed in \Cref{fig:categorical_methods}. A detailed description of the data formatting choices for categorical variable imputation can be found in the supplementary material in Section~\ref{sec:cat_imputation}. If a data set contains numerical and categorical variables, we refer to it as \emph{mixed data}.

\paragraph{Validation} During our simulation study, we systematically monitor and validate the outputs of each imputation method across all scenarios. This allows us to identify several types of imputation failures that compromise the validity of the results. The key error types are:

\begin{itemize}
    \item \textbf{Modification of observed values:} the imputation method altered originally observed values, which violates the core expectation that imputation should only affect missing entries.
    \item \textbf{Missing values remained after imputation:} the output still contains missing values, meaning the method failed to impute all missing data.
    \item \textbf{Computational error:} a broad category of failures during execution, including convergence errors, numerical instability (e.g., division by zero), or internal software errors, which caused the imputation method to crash. These typically interrupt the process and return a formal error message.
    \item \textbf{Timeout:}  the imputation method does not finish within the allocated time limit (10 hours for one imputation on a given data set). This time limit is intended to be a reasonable upper bound within which all methods should have imputed a data set. Although this may not be a formal algorithmic error, we cannot rule out the possibility that another type of error (e.g., convergence failure) would have occurred if more time had been allowed. Therefore, we treat timeouts the same way as any other error in our analysis. 
    \item \textbf{Incorrect labels in categorical variables} In the case of categorical data imputation, the method produced output outside the defined set of categories $\G$.
\end{itemize}
Importantly, each method that failed for computational reasons was given a second chance to complete the imputation, in order to account for occasional random failures (e.g., due to numerical instability or initialization issues) and give the method a fair opportunity to succeed.

\paragraph{Data sets} We evaluate the imputation performance on \nrofdatasets different data sets. These data sets span a variety of sizes (from $p \times n=$ 1080 to 2.9 million elements) and the four possible categories (complete or incomplete, numerical or mixed). \Cref{tab:datasets} displays all data sets with their origin and some statistics. 


\begin{table}[ht]
\centering
\begingroup\scriptsize
\begin{tabular}{lrccl}
  \hline
Name & Rows & Col. (Num./Cat.) & $\%$/Max Missing. & Source \\ 
  \hline
airfoil-self-noise & 1503 & 6 / 0 & 0 / 0 &  \citet{airfoil_self_noise_291} \\ 
  allergens & 2351 & 112 / 0 & 0 / 0 &  \citetalias{allergens_datachallenge} \\ 
  concrete & 1030 & 9 / 0 & 0 / 0 &  \citet{concrete_compressive_strength_165} \\ 
  enb & 768 & 10 / 0 & 0 / 0 &  \citet{energy_efficiency_242} \\ 
  fat & 252 & 18 / 0 & 0 / 0 &  \citet{R_faraway} \\ 
  scm1d & 9803 & 296 / 0 & 0 / 0 &  \citet{OpenML2013} \\ 
  scm20d & 8966 & 77 / 0 & 0 / 0 &  \citet{OpenML2013} \\ 
  windspeed & 433 & 6 / 0 & 0 / 0 &  \citet{R_mice} \\ 
  yeast & 1484 & 8 / 0 & 0 / 0 &  \citet{yeast_110} \\
  \hline
  choccake & 270 & 2 / 2 & 0 / 0 &  \citet{R_faraway} \\ 
  diamond & 53940 & 7 / 3 & 0 / 0 &  \citet{OpenML2013} \\ 
  electricity & 38474 & 7 / 2 & 0 / 0 &  \citet{OpenML2013} \\ 
  eye-movement & 7608 & 16 / 6 & 0 / 0 &  \citet{OpenML2013} \\ 
  german & 1000 & 2 / 19 & 0 / 0 &  \citet{south_german_credit_522} \\ 
  nels88 & 260 & 2 / 3 & 0 / 0 &  \citet{R_faraway} \\ 
  PimaIndiansDiabetes & 768 & 8 / 1 & 0 / 0 &  \citet{R_mlbench} \\ 
  worldcup & 595 & 5 / 2 & 0 / 0 &  \citet{R_faraway} \\ 
  \hline
    diabetes & 768 & 9 / 0 & 9.43 / 48.7 & \href{https://github.com/YBI-Foundation/Dataset}{YBI-Foundation/Dataset} \\ 
  globwarm & 1001 & 10 / 0 & 8.55 / 85.51 &  \citet{R_faraway} \\ 
  oceanbuoys & 736 & 8 / 0 & 3.01 / 12.64 &  \citet{R_naniar} \\ 
  popmis & 2000 & 6 / 0 & 7.07 / 42.4 &  \citet{R_mice} \\ 
  pulplignin & 301 & 22 / 0 & 5.32 / 46.84 &  \citet{R_VIM} \\ 
  \hline
  boys & 748 & 5 / 3 & 26.75 / 69.79 & \citet{R_mice} \\
  colic & 300 & 7 / 17 & 22.94 / 82.33  & \citet{R_VIM}\\
  debt & 464 & 3 / 10 & 4.20 / 13.36  & \citet{R_faraway}\\
  housevotes84 & 435 & 0 / 17 & 5.30 /  23.91 & \citet{R_mlbench}\\
  selfreport & 2060 & 5 / 5 &  20.00 / 61.02 & \citet{R_mice}\\
  soybean & 683 & 0 / 34 &  9.50 / 17.72 & \citet{R_mlbench}\\
  tbc & 3951 & 6 / 4 & 9.53 / 61.12 & \citet{R_mice} \\ 
  vnf & 1232 &  0 / 14 & 9.07 / 31.90 & \citet{R_missMDA} \\
  walking & 890 & 1 / 4 & 13.62 / 34.38 &  \citet{R_mice} \\ 
   \hline
\end{tabular}
\endgroup
\caption{Data sets used in the paper. For each data set, we report the number of rows, the number of numerical and categorical columns, the total proportion of missing values, as well as the maximum proportion of missing values found in any single column (in \%).} 
\label{tab:datasets}
\end{table}

\paragraph{Evaluation of Imputation for Artificially Created Missingness} As described previously, we will evaluate the performance of imputation methods according to the distributional distance of the imputed data set to the original data. To measure this distance when the complete data are available, we will consider the energy distance \citep{EnergyDistance} between the imputed and original data set (we provide more details in Appendix \ref{Evaluationmetricsdetails}). Although any other distributional distance could have been chosen, the energy distance has several advantages: It is easily and quickly estimated, without the need for specialized software (such as the Wasserstein distance \citep{villani2009wasserstein}, which is comparatively expensive to compute); it is theoretically sound and can be proven to be a proper distance between two distributions, as long as the first moment exists; moreover, no kernel or parameter needs to be chosen for the method to work well. As such \textit{``Energy Distance stands out among metrics for measuring distributional differences  due to its computational simplicity and broad applicability. Unlike Maximum Mean  Discrepancy (MMD), which relies on kernel selection in reproducing kernel Hilbert  spaces, or Wasserstein Distance, which is computationally intensive, Energy Distance  avoids these complexities while capturing both location and scale differences.'' }\citep{fan2025measuringheterogeneitymachinelearning}.


To apply the energy distance, we standardize each feature: the imputed and original values are scaled based on the mean and standard deviation from the original data sets, maintaining equal contribution of each feature. We also transform categorical variables into one-hot encoded variables. 

\paragraph{Evaluation of Imputation for Real Missingness} When the complete data are unavailable due to real missingness, we adopt the framework of proper Imputation Scores (I-Scores) introduced in \citet{ImputationScores}. Specifically, we use the energy-I-Score proposed in \citet{näf2025Iscore}. Another possible approach consists of leveraging an estimator of the Bures-Wasserstein criterion in the presence of missing data \citep{bleistein2025optimal}. This method assumes MCAR missingness and, given our interest in general-purpose imputation beyond this strict assumption, we prefer the energy-I-Score framework. The energy-I-Score can also treat categorical data by leveraging one-hot encoding. The number of internal imputations required by the energy-I-Score calculation was set to 20 throughout. We note that the energy-I-Score does not use any scaling of the data, as this may induce a bias when mean and variance cannot be estimated from a complete data set. We refer to Appendix \ref{Evaluationmetricsdetails} for more details. 

\paragraph{Ranking the Methods} Given the wide variation in data set sizes and dimensions—which influences the absolute value of the energy distance and energy-I-Score— and the necessity to handle unsuccessful imputations, we evaluate methods not directly by their energy distance/energy-I-Score, but based on their \emph{rank} according to these metrics. That is, for each given data set, missing mechanism,  missingness proportion and replicate, we rank the methods from smallest to largest (standardized) value of energy distance/energy-I-Score. For more details on the calculation of these metrics, we refer to Appendix \ref{Evaluationmetricsdetails}. Methods that fail for any reason are assigned a rank one position worse than the lowest-ranked successful method.  For example, if 58 out of 60 methods successfully impute the data and 2 fail according to our criteria, those failing methods receive rank 59. We then generate a final ranking by averaging each method's rank across all setups (i.e., data sets, missing mechanisms, missingness proportions and replicates). This approach ensures fair weighting across diverse data sets and appropriately penalizes methods that fail to complete.

\section{Summary of Findings and Limitations}\label{Sec_Discussion}

Before presenting our results in detail in Sections \ref{Sec_ArtificialMissing} and \ref{Sec_Realmissing}, we summarize our findings and recommendations in this section. All the methods discussed here with their references and implementations can be found in Table \ref{tab:methods}.

The goal of this paper is to reduce the large number of missing value imputation methods to a handful. A first important caveat is that this is not as straightforward as in the case of prediction. Even if one agrees that preserving the distribution of the data is the main goal, measuring distributional distances is not trivial. Moreover, the true missing value mechanism is generally unknown. Nevertheless, for the considered data sets we see a remarkably consistent picture, when using artificial MCAR/MAR mechanisms and the energy distance as well as when using real missing data sets and the energy-I-Score. Overall, FCS methods outperform joint methods, in particular mice methods tend to have the best performance. Despite their growing popularity, deep learning-based joint imputation methods, such as \texttt{miwae}, \texttt{gain}, \texttt{miracle}, and \texttt{remasker} consistently underperformed in our analysis. This suggests that, despite potentially higher computational cost, iterative approach offers a substantial benefit in practical imputation performance. In particular, we find consistently that: 
\begin{itemize}
    \item For purely numerical data, in terms of energy distance $\&$ artificial missingness, \texttt{mice\_cart}, \texttt{aregImpute}, \texttt{mice\_rf} and \texttt{hyperimpute} 
    stand out. In addition, \texttt{mice\_mixed} performs very well, although its performance is concealed by imputation and timeout errors that are heavily penalized.
    \item For mixed data, in terms of energy distance $\&$ artificial missingness, \texttt{mice\_cart}, \texttt{mice\_mixed}, \texttt{mice\_pmm} stand out, while in terms of energy-I-Score $\&$ real missingness the same is true with \texttt{mice\_cart} replaced by \texttt{mice\_rf} (\texttt{mice\_rf}, \texttt{mice\_mixed}, \texttt{mice\_pmm}).
\end{itemize}

We note that \texttt{mice\_rf} is a somewhat curious case: It performs well for numerical data with artificial missingness and replaces \texttt{mice\_cart} as the best method in the case of real missingness, using the energy-I-Score. At the same time, it performs quite poorly for mixed data with artificial missingness, with no obvious explanation. Moreover, \texttt{mice\_mixed} corresponds to \texttt{mice\_default} with the number of iterations increased from the default five to one hundred, which explains the relatively large number of timeouts despite the moderate dimensionality of the data sets. We also verified that increasing the number of iterations for other top-performing methods, such as \texttt{mice\_cart}, can further improve their performance beyond the best results reported here. However, these extended versions are not included in the benchmark, as our goal was to evaluate methods under their default configurations.


\paragraph{Recommendation} Our recommendation for general-purpose i.i.d. mixed-data imputation would thus be to consider the FCS methods \texttt{mice\_cart}, \texttt{mice\_pmm}, \texttt{mice\_rf} and potentially \texttt{hyperimpute}. We note that we exclude \texttt{mice\_mixed} from our recommendation, as there are methods which are as good using five instead of one hundred iterations. These methods also stand out in terms of numerical stability, with almost no errors or convergence issues. It appears that the only real downside to FCS methods would be their computational time, especially for larger data sets. On the other hand, for the relatively low-dimensional data sets considered in this paper, we found no obvious relationship between the computational complexity of a method (measured by runtime) and its imputation quality, except for the consistently poor performance of naive approaches. Moreover, it should be mentioned here that this is also an issue of implementation; the \texttt{mice} package \citep{mice2023} is written in base \textsf{R}, making iterations over columns relatively slow. 

\paragraph{Limitations} We note some limitations of the current study: First, while we included data sets with a high number of observations (up to 53'940), our data sets are relatively low-dimensional; at most 296 columns, most smaller. This is due to the fact that preference was given to data sets that have been used in previous benchmarking studies or published analyses. Results might be different for higher dimensional data sets, in particular, it is likely that the imputation times for the FCS methods explode. We also did not consider data sets with less than 200 observations, as for these datas ets even the estimation of the energy distance might become unreliable. Second, we treat all data sets as i.i.d. and do not consider data sets with obvious time-series or panel data structure. Third, we note that such a comparison of methods is necessarily a comparison of \emph{implementations} of methods. For instance, while \texttt{hyperimpute} does the same $d$ FCS iterations to impute in Python, it tends to be much faster than the mice methods implemented in base \textsf{R} in the \texttt{mice} package. At the same time, the \texttt{mice} \textsf{R} package has been widely used and has gone through several iterations, making the methods implemented within the package very stable. This may be in contrast to more recent methods, such as \texttt{mice$\_$drf}, which tends to have more numerical errors in our analysis, despite being a mice method as well.



\section{Artificial Missing Value Mechanisms}\label{Sec_ArtificialMissing}

We start by artificially generating missing value mechanisms as described in Section \ref{Sec_Setup}. In the first step, we consider data sets consisting only of numerical variables, that is, random variables that (presumably) arise from continuous distributions. In the second step, we consider mixed data sets that also include categorical and ordinal variables.

Previous studies have often focused solely on numerical variables. In fact, according to our definition, only \nrofmethodscat out of \nrofmethods methods are able to impute categorical data. Among these, \nrofmethodboth can handle mixed-type data sets, containing both continuous and categorical variables. In contrast, \texttt{missmda\_famd\_em} and \texttt{missmda\_famd\_reg} are restricted to categorical data and are not applicable when only continuous variables are present. We also note the case of \texttt{mice\_default}, which behaves identically to \texttt{mice\_pmm} when imputing numerical data, but employs a different regression model for categorical variables. For this reason, \texttt{mice\_default} is evaluated only in the mixed-data setting.


\subsection{Numerical Data}

\begin{figure}
    \centering
    \includegraphics[width=01 \linewidth]{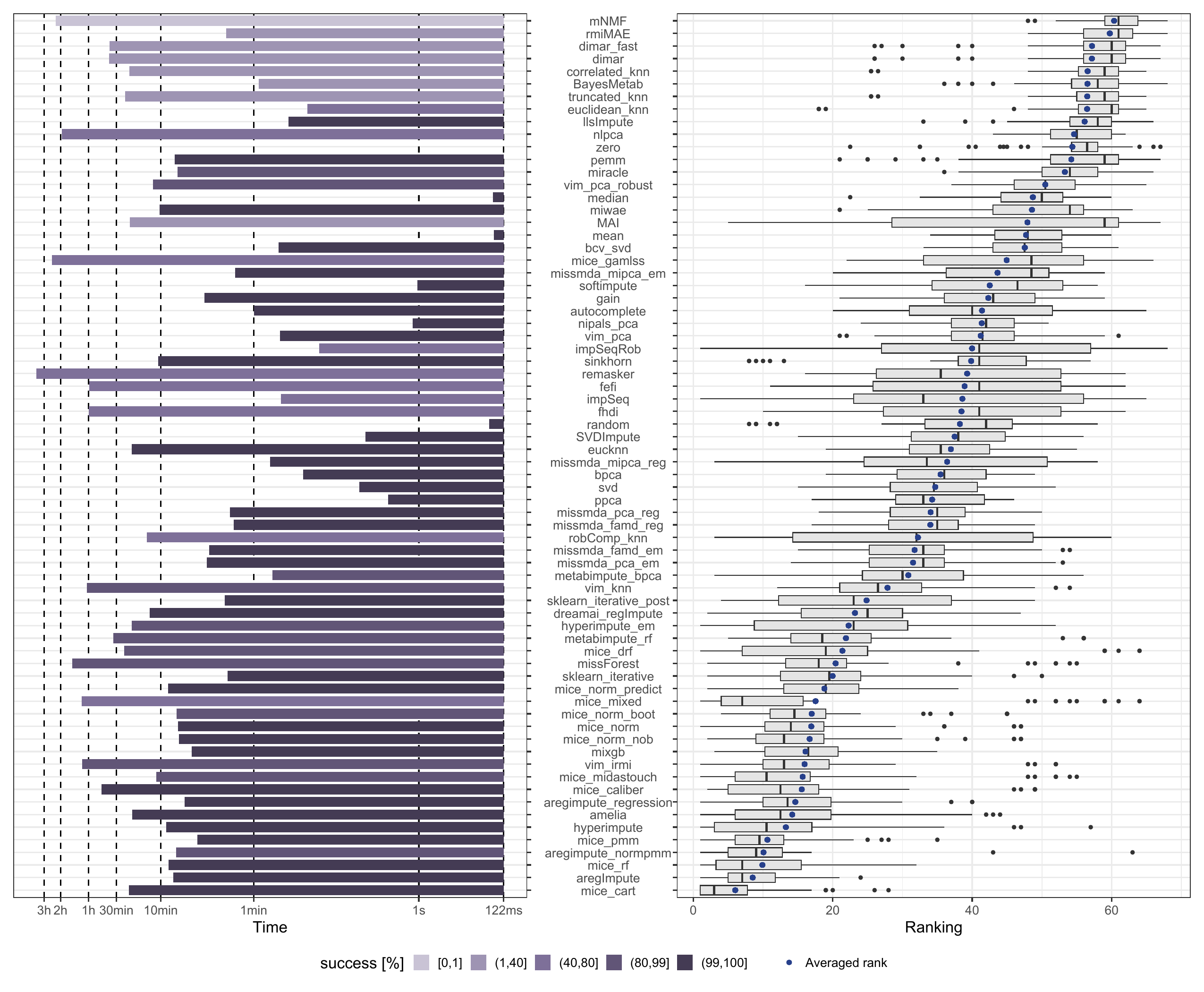}
    \caption{Results for artificial missingness, on numerical data sets. Left: averaged evaluation time. Right: position in ranking averaged over 2 replications of amputation.}
    \label{fig:FakeMissing_numericalData}
\end{figure}

Figure~\ref{fig:FakeMissing_numericalData} summarizes the results for numerical data sets, displaying the performance of \nrofmethodsnum methods capable of imputing numerical variables. It presents a boxplot of method ranks across all combinations of data sets, missingness ratios, and missingness mechanisms (with replicates), as well as the average runtime for each method.

Building on these results, our analysis of complete numerical data sets revealed several key conclusions. 
We observe that FCS imputation methods, such as \texttt{mice\_cart}, \texttt{mice\_rf}, and \texttt{aregImpute}, consistently dominate the rankings. In fact, all of the top five methods across numerical data sets employ FCS imputation. This suggests that, despite potentially higher computational cost, the FCS approach offers a substantial benefit in practical imputation performance. Among the evaluated methods, \texttt{mice\_cart} consistently emerges as the top performer. In particular, it achieves a median ranking markedly better than other methods (median rank = 4). In contrast, despite their computational complexity and growing popularity \citep[e.g., ][cited over 1,500 times]{GAIN}, deep learning-based joint imputation methods, such as \texttt{miwae}, \texttt{gain}, \texttt{miracle}, and \texttt{remasker} consistently underperform in our analysis. Furthermore, our evaluation reveal some surprising findings. The distributional imputation method \texttt{sklearn\_iterative\_post} tends to perform worse than its non-distributional variant \texttt{sklearn\_iterative}. Similarly, despite their non-distributional nature, methods such as \texttt{missForest} perform remarkably well, indicating that some conditional mean imputation approaches can still recover the distribution well in practice. However, we note that this also reflects our decision to standardize the data before applying the energy distance. Without this standardization, \texttt{missForest} falls further behind, while methods like \texttt{mice\_cart} keep their favorable ranking, as seen in Figure \ref{fig:measures}.  We note that \texttt{mice\_mixed} performs relatively poorly only because the average ranking is strongly influenced by cases where the method resulted in numerical or timeout errors. If we had decided to give these cases less influence by considering the median ranking instead, the method would have ranked among the best five methods, as seen in Figure \ref{fig:FakeMissing_numericalData_median}.


Overall,  we find that a small set of complementary methods is often sufficient to ensure that at least one ranks among the top k performers across a wide range of scenarios. For example, as shown in Figure~\ref{fig:top-k}, using only four distinct methods (\texttt{mice\_cart},
\texttt{hyperimpute},
\texttt{aregImpute},
\texttt{hyperimpute\_em}) provides over 90\% coverage of top-3 rankings. This highlights a strong potential to reduce the full set of \nrofmethodsnum methods to a much smaller, well-curated subset without compromising imputation performance across diverse settings.

\begin{figure}
    \centering
    \includegraphics[width=1\linewidth]{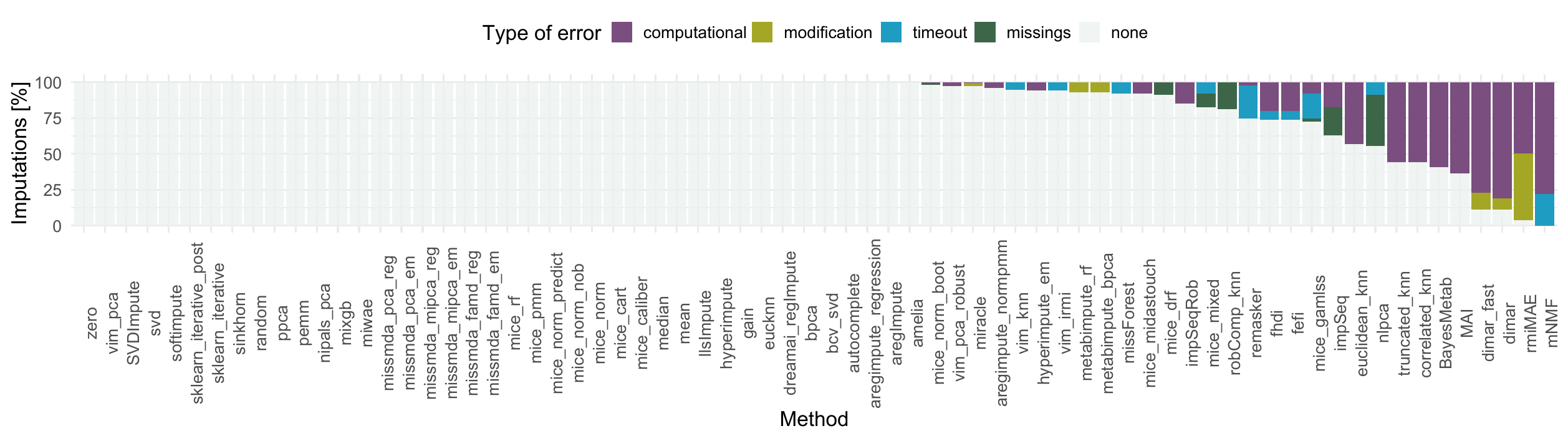}
    \caption{\textit{Fractions of different types of errors obtained by all the methods on numeric data sets with artificial missingness.}}
    \label{fig:errors_num}
\end{figure}

A significant finding regarding stability was that almost half ($30$ out of $\nrofmethodsnum$) of the methods encountered computational errors or failed to complete within the set time limits, indicating substantial stability issues (Figure \ref{fig:errors_num}). Overall, we observe that the less stable methods tend to rank lower, which is a direct result of the ranking strategy we adopted. Across all scenarios, approximately $11.85\%$ of imputations resulted in some kind of error. The most common failure type was computational errors (e.g., crashes, convergence problems), which accounted for $8.1\%$ of all imputations. Other error types included unexpected presence of missing values in the output ($1.27\%$) and unexpected modifications to the original data ($1.02\%$). The cases where the method exceeded 10 hours  were relatively rare, representing only $1.47\%$ of all runs. Notably, $88.1\%$ of imputations completed successfully without any issues. In $35$ cases, a failed imputation was successfully completed on the second attempt. These involved $6$ different methods (\texttt{amelia}, \texttt{aregimpute\_normpmm}, \texttt{hyperimpute\_em}, \texttt{mice\_norm\_boot}, \texttt{mice\_midastouch}, \texttt{miwae}) that initially returned an error but managed to produce valid imputations when given another chance. The majority of computational errors and timeouts were observed on the largest data sets, while other types of errors appeared to be relatively evenly distributed (see Figure \ref{fig:errors_datasets} in Appendix \ref{Sec_Stability}). Again, the methods based on the \texttt{mice} framework, in particular \texttt{mice\_cart} had relatively little stability issues, performing consistently well across the tested scenarios. 

Interestingly, we found no obvious relationship between the computational complexity of a method (measured by runtime) and its imputation quality, except for the consistently poor performance of naive approaches. Moreover, we observed no substantial differences in runtime across missingness mechanisms (MCAR, MAR) or missingness ratios, as presented in Figure~\ref{fig:times_cases}. We also did not observe any clear link between a method’s runtime and its stability, suggesting that longer computation does not necessarily lead to more robust execution and conversely, unstable methods were not always the slowest, as shown in Figure~\ref{fig:times_success} in the appendix.

To obtain a synthetic view of the relationships between imputation methods across different simulation scenarios, we performed a principal component analysis (PCA). In this analysis, the variables correspond to specific simulation cases, defined by combinations of data sets, missingness patterns, and missing data proportions. The observations are imputation methods, with input values given by their ranking positions based on the standardized energy distance.

\begin{figure}
    \centering
    \includegraphics[width=1\linewidth]{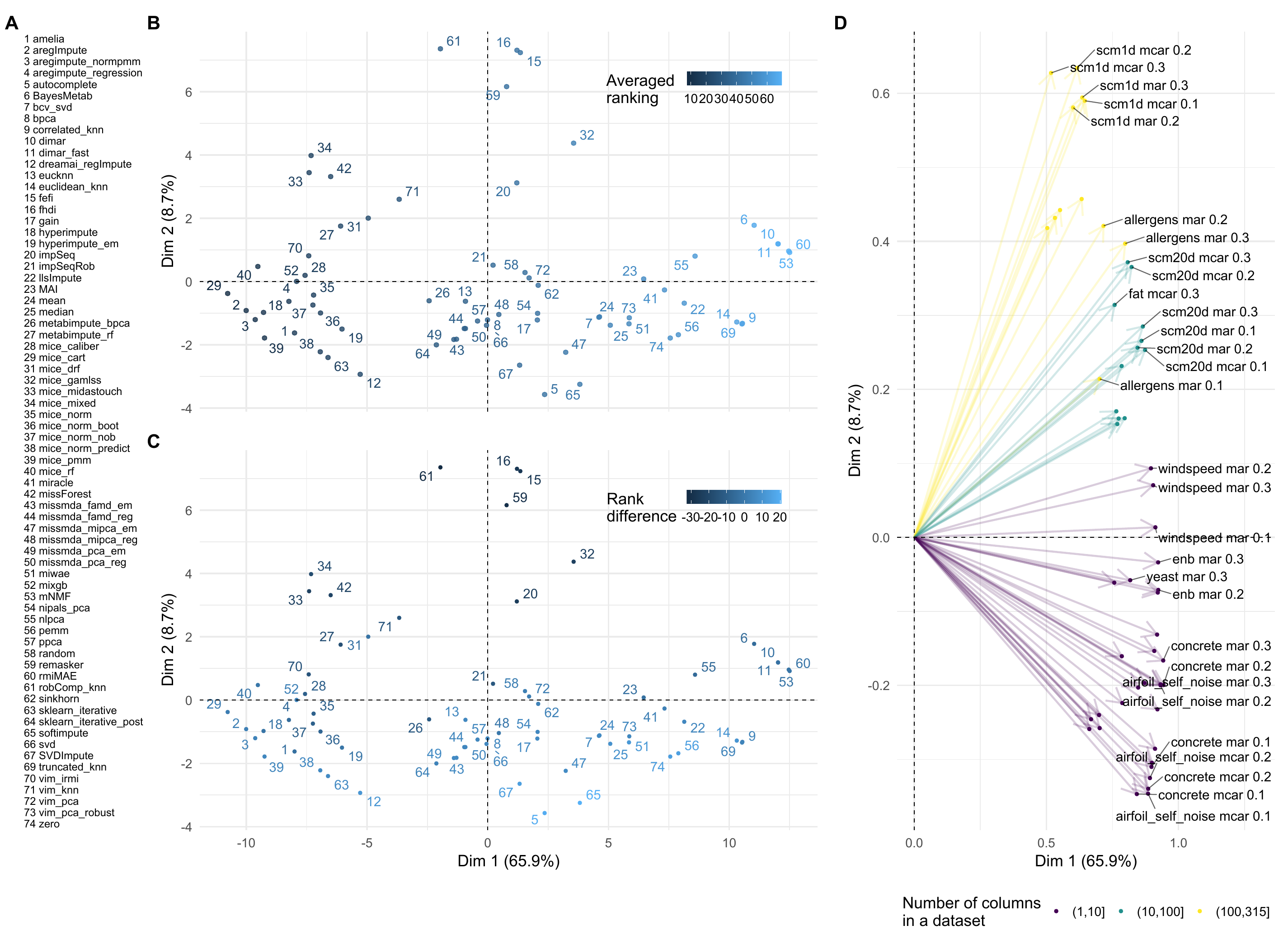}
    \caption{\textit{Principal Component Analysis of imputation method rankings across simulation scenarios.
    (A) All methods in alphabetical order
(B) Methods plotted in the space of the first two principal components; color indicates the average rank across all simulation scenarios (lower is better); number corresponds to A.
(C) Same projection as in A, but color represents the difference between the average rank and the average rank on large data sets (positive values indicate better performance on large data sets, negative values indicate better performance on small data sets, values near zero denote stable performance regardless of data set size).
(D) Biplot of all simulation scenarios, with color indicating data set size.}}
    \label{fig:PCA}
\end{figure}

The first principal component explains the largest proportion of total variance 65.9\%, while the second accounts for 8.7\%. We interpret the first component as the main axis of imputation quality - methods with lower scores along this dimension tended to achieve better average ranks in all scenarios (\Cref{fig:PCA} B.). In particular, \texttt{mice\_cart}, \texttt{areg}-based methods, \texttt{mice\_pmm}, and \texttt{hyperimpute} occupy the upper positions. \Cref{fig:PCA} D. shows that the first component correlates most strongly with simulation cases involving small and medium-sized data sets (up to 100 variables). In contrast, the second component is primarily associated with cases involving large data sets (more than 100 columns). As shown in Figure \ref{fig:PCA} C., this component differentiates the methods according to their performance depending on the size of the data set. Notably, for instance \texttt{remasker} and \texttt{mice\_mixed} perform substantially better on smaller data sets, whereas \texttt{autocomplete} or \texttt{softimpute} achieve relatively stronger rankings in large-data scenarios.

In summary, we observe that the first principal component is not dominated by any single simulation case, indicating a well-balanced benchmark design. No specific data set, missingness pattern, or missingness proportion appears to have disproportionately influenced the overall ranking structure. Furthermore, we did not observe substantial differences in performance between 10\%, 20\%, and 30\% missing data, nor between MAR and MCAR scenarios which suggests a relative stability of methods across varying degrees and mechanisms of missingness.

\subsection{Mixed Data}

\begin{figure}
    \centering
    \includegraphics[width=1 \linewidth]{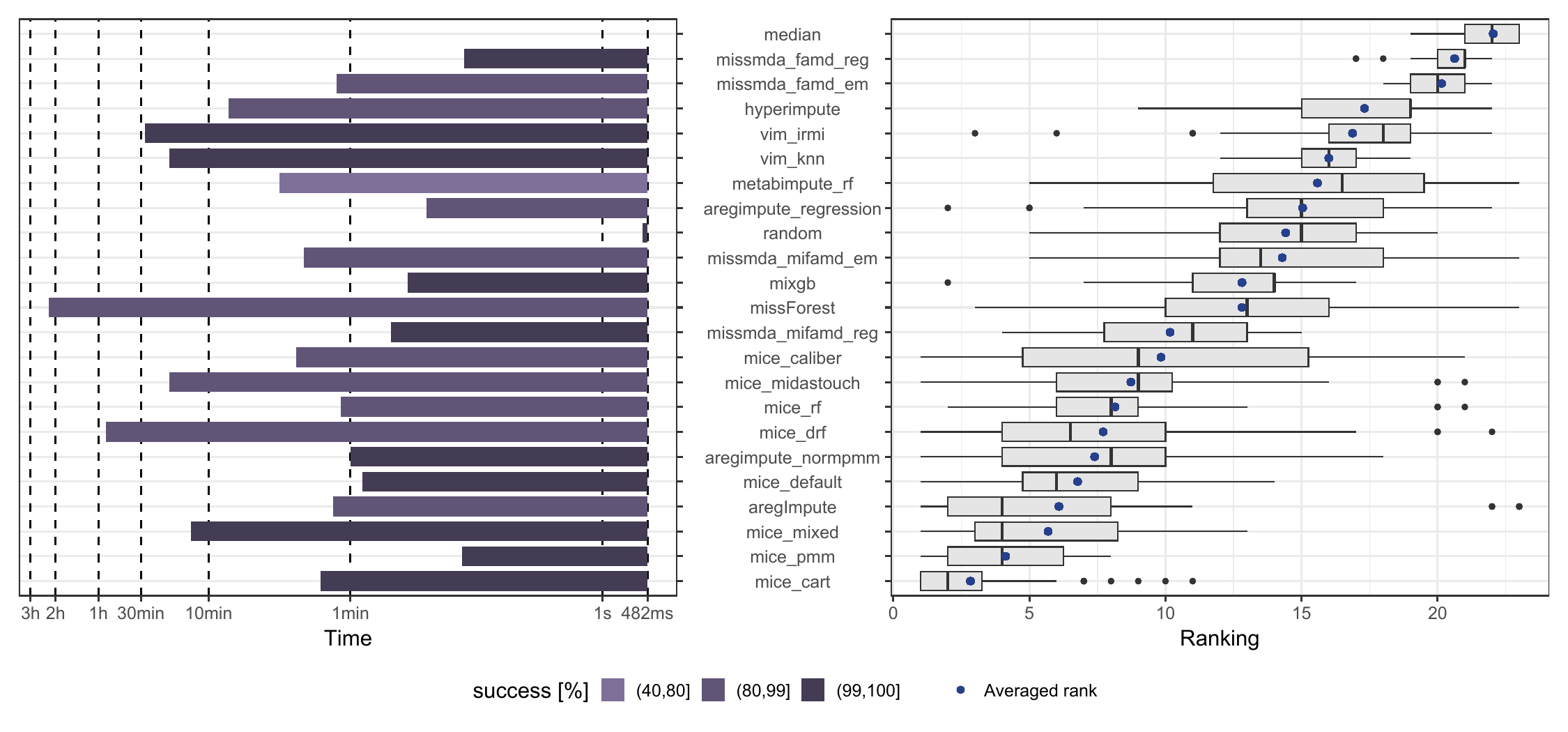}
    \caption{Results for artificial missingness, on mixed data sets. Left: averaged evaluation time. Right: position in ranking averaged over 2 replications of amputation.}
    \label{fig:FakeMissing_mixedData}
\end{figure}

Having purely numerical data appears somewhat unrealistic in modern applications. As such, we will now repeat our analysis with data sets that contain at least one column of categorical variables, referred to as mixed data. The number of methods that can deal with mixed data is markedly reduced, leading to a much smaller set of imputations. The results are displayed in \Cref{fig:FakeMissing_mixedData}. Again, \texttt{mice\_cart} takes the lead here, followed by the set of other FCS methods such as \texttt{areg\_normpmm}, \texttt{mice\_mixed}, \texttt{mice\_default}, \texttt{mice\_pmm} and \texttt{mice\_drf}. 
Surprisingly, despite its strong performance on numerical data sets, \texttt{hyperimpute} demonstrates noticeably weaker results when applied to categorical data. Furthermore, it is interesting to compare the performance of \texttt{random} and \texttt{median} imputations. In the case of categorical variables, the \texttt{median} imputation reduces to mode imputation, as categories lack a natural ordering. Thus, it corresponds to assigning the most frequent category to all missing values in a given feature. This results in a constant imputed value per feature, distorting the original distribution and leading to poor performance in our analysis. In contrast, \texttt{random} imputation introduces variability, occasionally selecting less frequent categories, which can better preserve marginal distributions. 

Building on the earlier findings for numerical data sets, we observe a similar pattern for mixed data. A small, complementary set of methods is sufficient to capture the vast majority of top rankings. In fact, just three methods (\texttt{mice\_cart}, \texttt{mice\_mixed}, \texttt{mice\_pmm}) achieve 90\% coverage of top-3 rankings (Figure \ref{fig:top-k_cat}).

\begin{figure}
    \centering
    \includegraphics[width=1\linewidth]{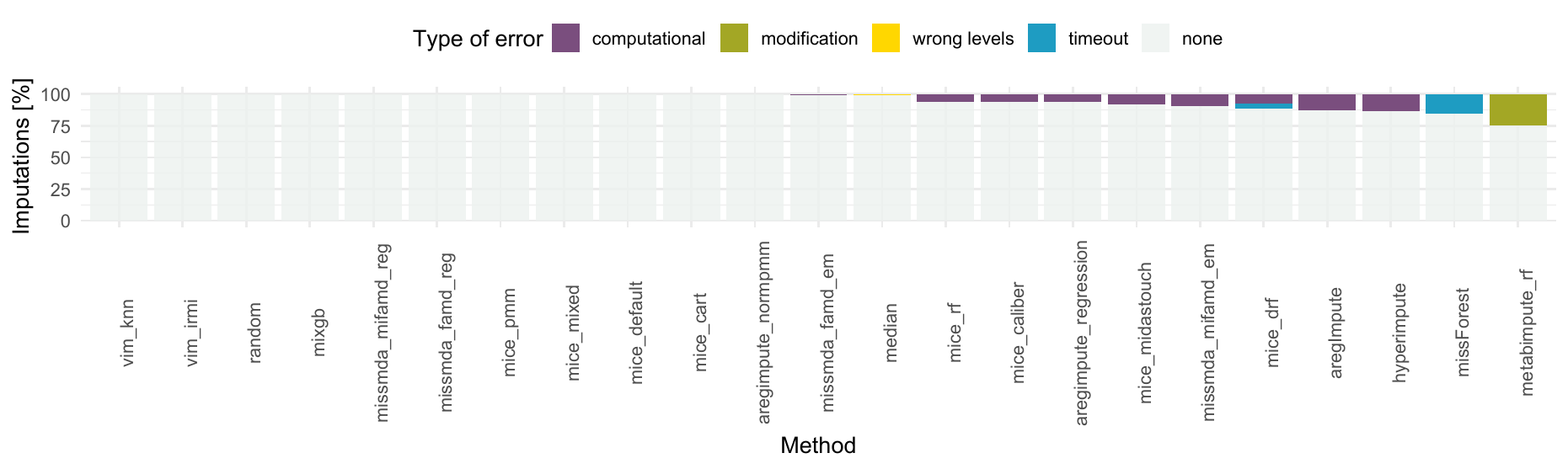}
    \caption{\textit{Fractions of different types of errors obtained by all the methods on mixed data sets with artificial missingness.}}
    \label{fig:errors_mixed}
\end{figure}

Moreover, we observe a high proportion of successful imputations ($94.93\%$), as shown in Figure \ref{fig:errors_mixed}. The most common errors were computational issues ($3.08\%$), followed by unintended modifications ($1.09\%$) and cases where the time limit was exceeded ($0.86\%$). The rarest problem involved imputations with incorrect categories ($0.05\%$). In 12 cases, the initial imputation attempt failed, but the method succeeded on the second attempt. These instances involved the following four  methods: \texttt{aregImpute}, \texttt{aregimpute\_regression}, \texttt{hyperimpute} and \texttt{missmda\_mifamd\_em}. Once again, we did not observe any relationship between computation time and imputation quality.


\begin{figure}[h]
    \centering
    \includegraphics[width=1 \linewidth]{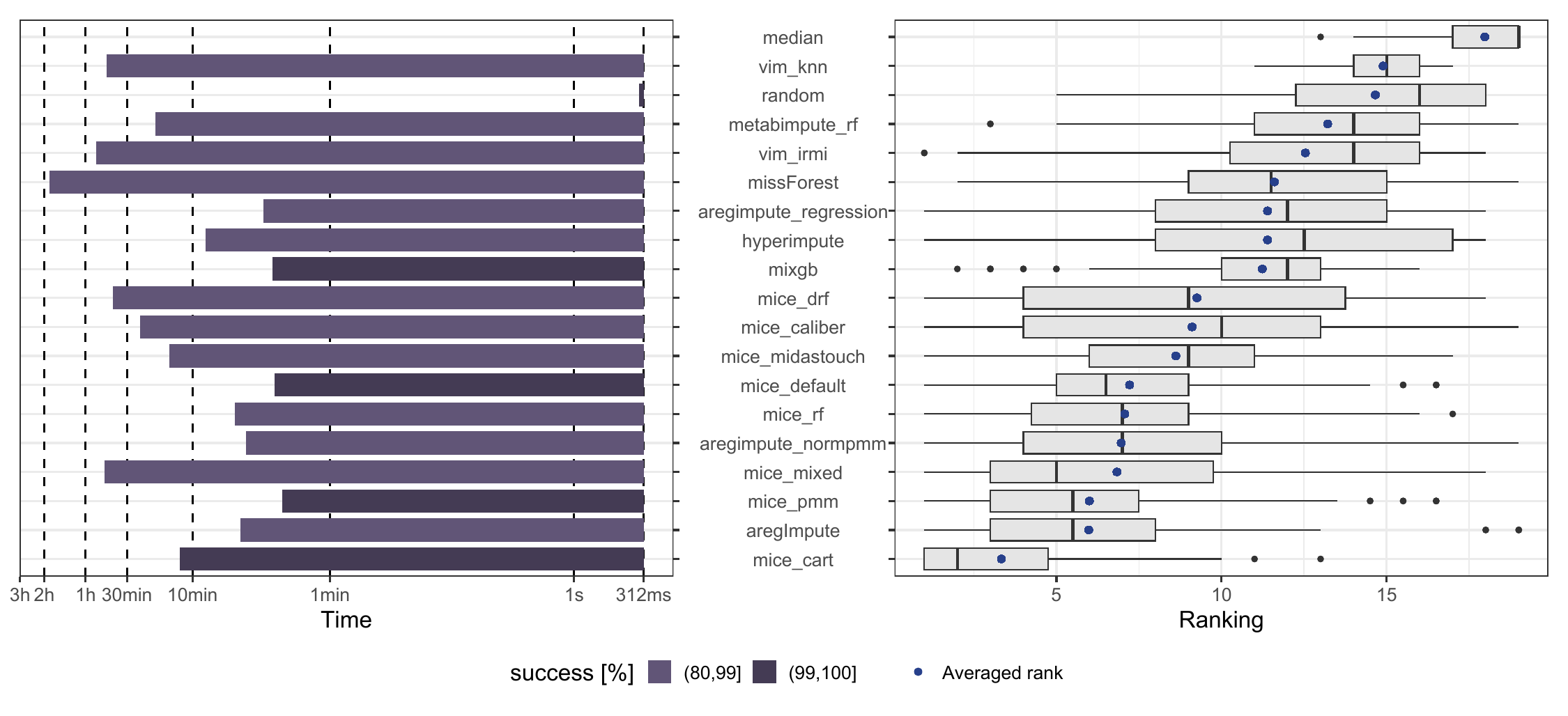}
    \caption{Results for artificial missingness, on numerical and mixed data sets. Left: averaged evaluation time. Right: position in ranking averaged over 2 replications of amputation.}
    \label{fig:FakeMissing_Overall}
\end{figure}

\Cref{fig:FakeMissing_Overall} aggregates the performance on both mixed-type and purely numerical data sets, considering only methods that successfully handled both data types. Once mixed-type scenarios were included in the analysis, \texttt{mice}-based methods occupied all leading positions, underscoring their robustness and versatility in handling heterogeneous data structures. Among them, \texttt{mice\_cart} emerges as the clear top performer in both numerical and mixed data sets, followed by \texttt{mice\_pmm} and \texttt{aregImpute}. This further highlights the consistent performance of FCS approaches across different data types. Finally, although this aggregated view offers a clear overall ranking, it may conceal performance differences across data types. For instance, some methods such as \texttt{hyperimpute} demonstrate strong performance on numerical data but perform poorly on categorical variables. Hence, the selection of an imputation method might benefit from considering the data type.


\begin{figure}
    \centering
    \includegraphics[width=1\linewidth]{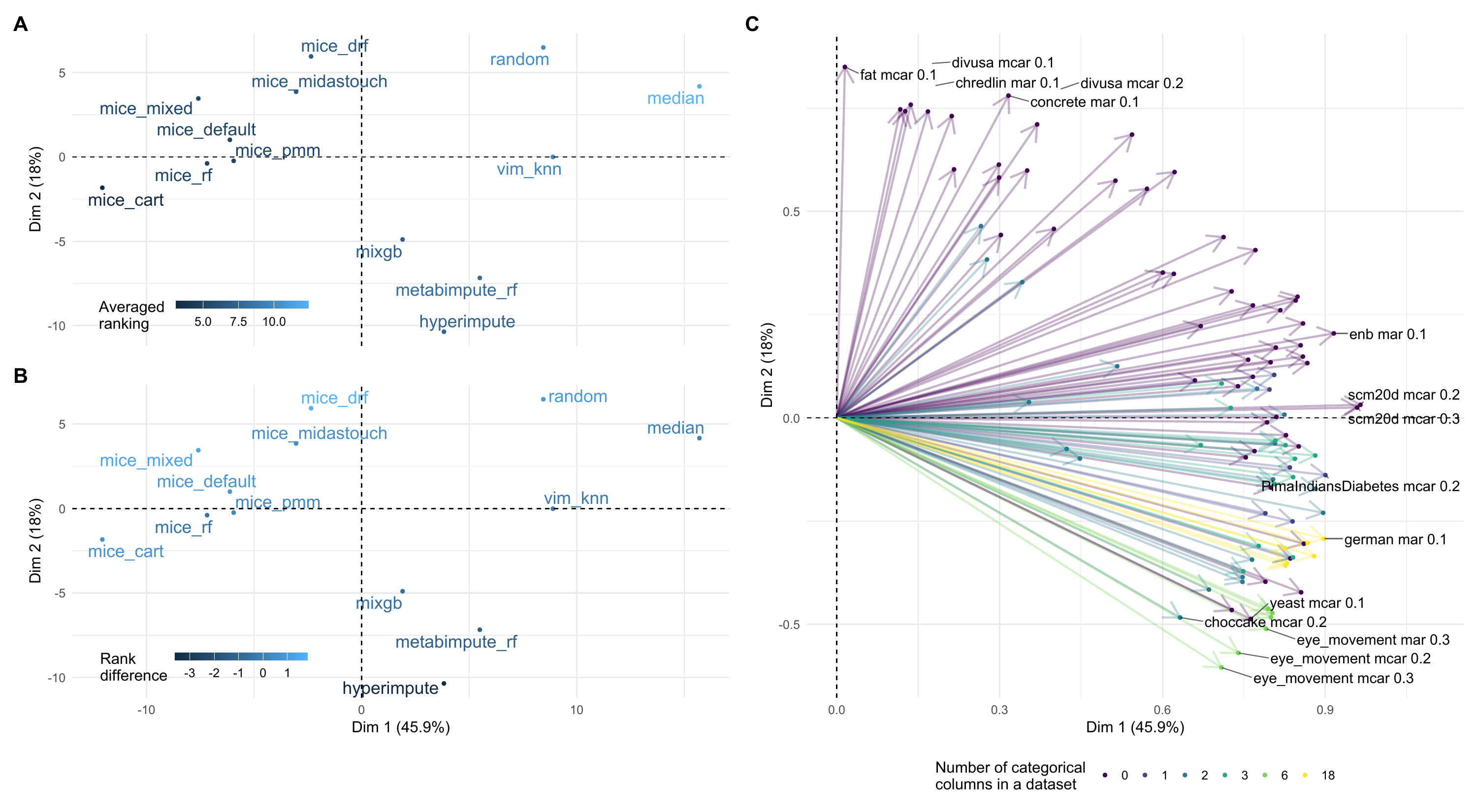}
    \caption{\textit{Principal Component Analysis of imputation method rankings across complete simulation scenarios (numerical and mixed data sets).
(A) Methods plotted in the space of the first two principal components; color indicates the average rank across all complete scenarios (lower is better).
(B) Same projection as in A, but color represents the difference between the rank averaged over all scenarios and the average rank on mixed data sets (positive values indicate better performance on mixed data, negative values indicate better performance on purely numerical data, values near zero denote stable performance across data types).
(C) Biplot of all simulation scenarios, with color indicating the number of categorical columns in a data set.}}
    \label{fig:PCA_mixed}
\end{figure}

Figure \ref{fig:PCA_mixed} shows the results of PCA performed jointly on mixed-type and numerical data sets. Once again, we observe that the first principal component, this time explaining over 45\% of the total variance, captures the overall performance of the methods. The second component (over 12\%), on the other hand, appears to represent performance on imputing categorical variables. Thus, in Figure \ref{fig:PCA_mixed} B. we can notice that  \texttt{mice\_cart} and \texttt{areg} methods demonstrated stable performance regardless of the number of categorical variables to be imputed. The second principal component also highlights the performance gap of \texttt{hyperimpute} between mixed-type and purely numerical data sets, suggesting that its performance is more sensitive to the presence of categorical variables. Finally, we do not observe any scenario that appears to disrupt the stability of our simulation.

\section{Real Missing Value Mechanisms}\label{Sec_Realmissing}

\begin{figure}
    \centering
    \includegraphics[width=1 \linewidth]{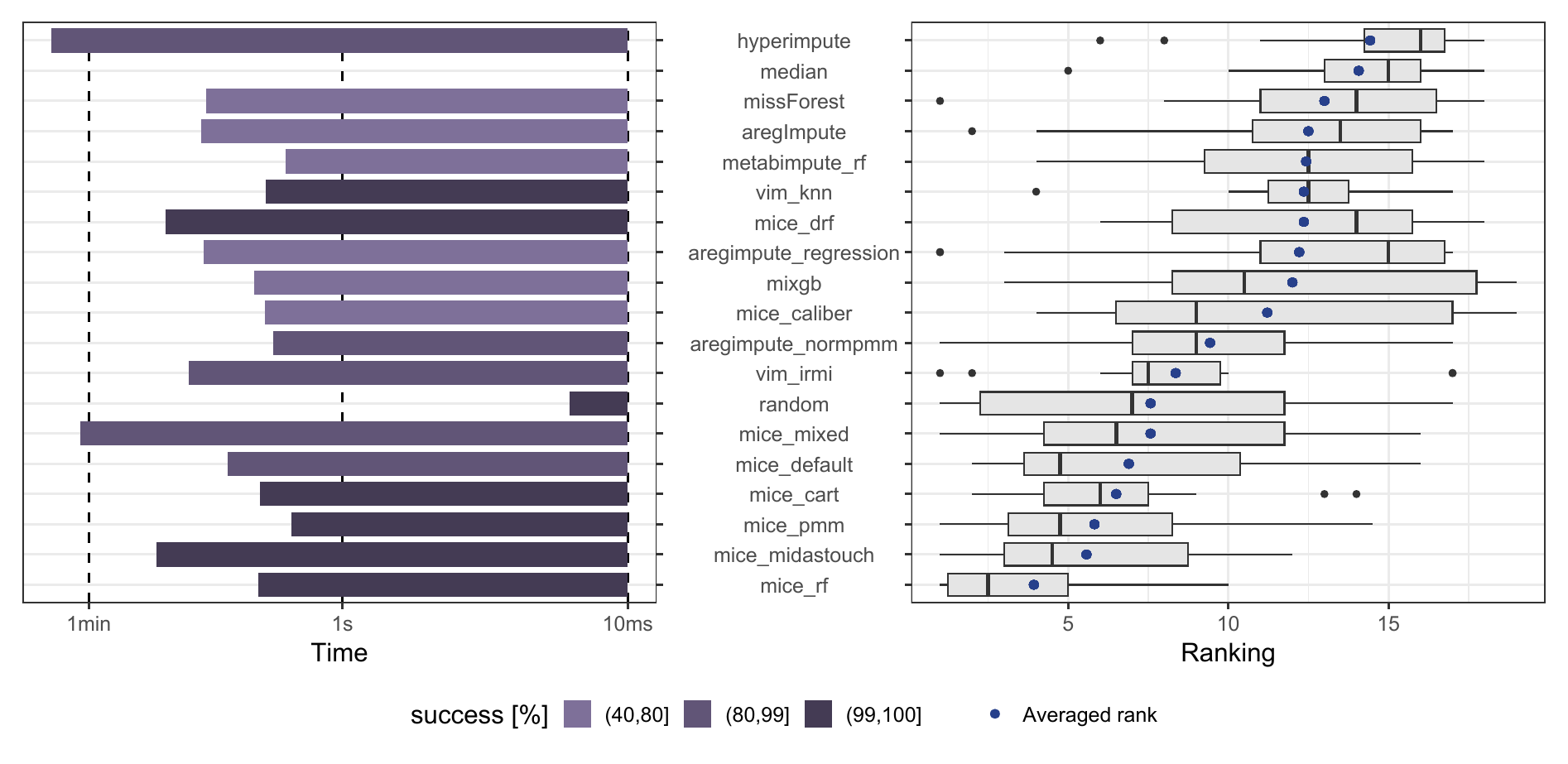}
    \caption{Results for real missingness, on numerical and mixed data sets. Left: averaged evaluation time. Right: position in ranking.}
    \label{fig:RealMissing}
\end{figure}

In this section, we evaluate the performance of imputation methods on data with real missingness. We thereby focus exclusively on methods capable of imputing mixed-type data and evaluate their performance on both numerical and mixed data sets. Since we assess the results jointly across numerical and mixed data sets, methods designed to handle only one of these data types are excluded from the analysis, resulting in a total of 19 evaluated methods. Figure \ref{fig:RealMissing} summarizes the results for numerical and mixed data sets. As in the previous analysis, we present method rankings, but this time based on the energy-I-Score, which directly reflects the quality of the imputation. It is important to note that the energy-I-Score involves repeated imputations across different subsets of columns and rows, and thus implicitly requires methods to be stable under minor data perturbations. Failures during score computation therefore indicate an inherent lack of robustness rather than a limitation of the evaluation procedure itself. Consequently, we classify such cases as computational errors, as these methods were unable to reliably complete the imputation process required for the energy-I-Score.

Despite the different metric (energy-I-Score instead of the energy distance) and the potentially different missing value mechanisms (real vs artificial), the dominance of the FCS approach is consistent. While \texttt{mice\_cart} leads the ranking in controlled missingness, on real missingness with numerical and mixed data sets, \texttt{mice\_rf} dominates. The reason for this is likely that the energy-I-Score is highly sensitive to the quality of the imputation draws, as discussed in \cite{näf2025Iscore}. In this context, \texttt{mice\_cart} tends to generate less diverse imputations due to its reliance on a single tree, resulting in small and discrete donor sets and an underestimation of uncertainty. By contrast, \texttt{mice\_rf} employs an ensemble of trees, producing a larger and more representative donor pool that better captures the underlying variability. Consequently, imputations obtained from \texttt{mice\_rf} tend to better reflect the underlying conditional distribution. While this might not be picked up by the energy score when observing a single data set, it translates into more realistic draws from the perspective of the energy-I-Score which relies on generating multiple imputations from the conditional distribution. This is likely also the reason why \texttt{missForest} ranks lower than when using the energy distance. Despite this, \texttt{mice\_cart} still scores high. Closely behind \texttt{mice\_rf} in the ranking are other FCS methods, including \texttt{mice\_midastouch}, \texttt{mice\_pmm}, \texttt{mice\_cart}, and \texttt{mice\_default}. Their strong and consistent performance across numerical and mixed data sets further illustrates the robustness of FCS strategies, extending beyond artificial missingness to more complex real-world missingness patterns. Moreover, \texttt{mice\_midastouch} ranks ahead of \texttt{mice\_pmm}, once again illustrating the sensitivity of the energy-I-Score to the quality of the imputation draws. Unlike standard predictive mean matching, which selects a donor uniformly at random from a fixed set of nearest neighbors, \texttt{mice\_midastouch} assigns higher probabilities to donors that are closer to the predicted value. This weighted donor selection leads to imputations that more accurately reflect the underlying conditional distribution, resulting in improved performance under the energy-I-Score. Similarly to the artificial missingness setting, \texttt{hyperimpute} performed poorly on mixed data sets, indicating again its relatively poor handling heterogeneous data types. Remarkably, only three algorithms (\texttt{mice\_rf}, \texttt{mice\_midastouch}, and \texttt{aregimpute\_normpmm}) are responsible for 100\% of appearances in the top four positions in all scenarios (Figure \ref{fig:top-k_inc}).

\begin{figure}
    \centering
    \includegraphics[width=1\linewidth]{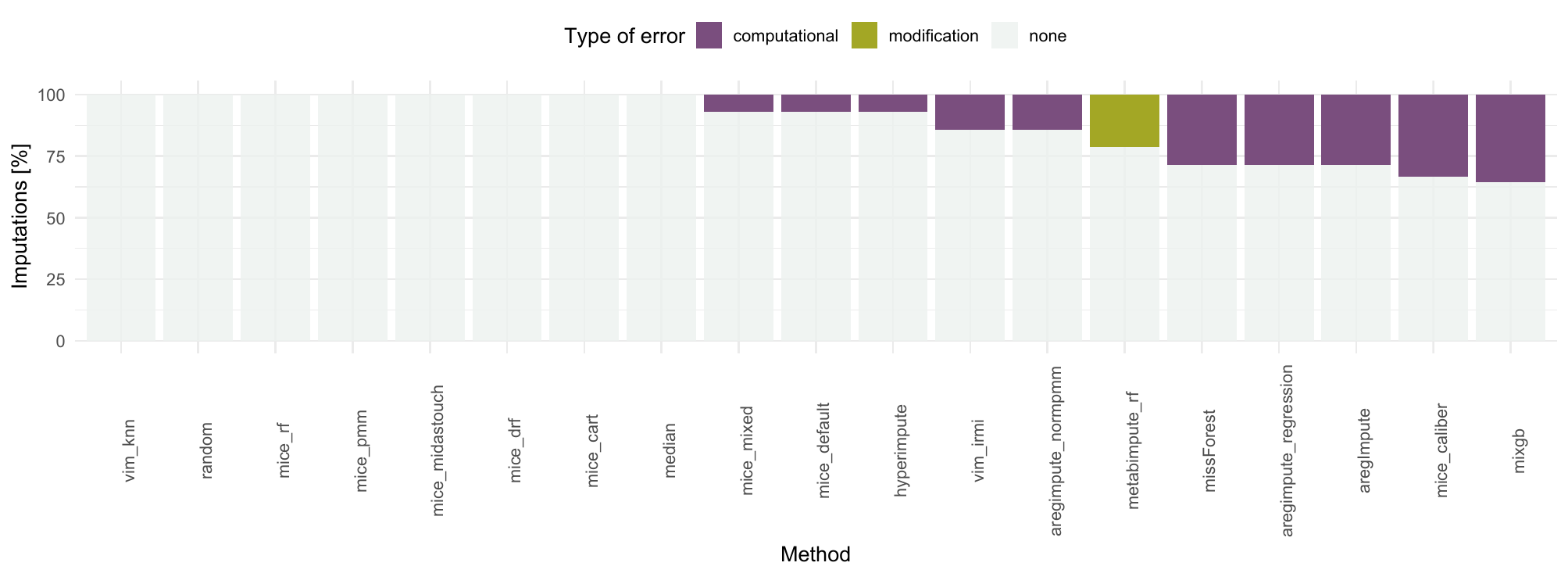}
    \caption{\textit{Fractions of different types of errors obtained by all the methods on mixed data sets with real missingness.}}
    \label{fig:errors_incomplete}
\end{figure}

Compared to the controlled missingness experiment, the overall error rate is notably higher (12\%), indicating that real-world missingness poses a greater challenge for many imputation algorithms (Figure \ref{fig:errors_incomplete}). This increased instability reflects the more complex and heterogeneous structure of real missingness patterns, which can exacerbate both convergence issues and sensitivity to data perturbations. The most common error type was computational errors (10.9\%) and the second most frequent category was unexpected modification errors (1.1\%). This time all the methods that managed to impute the data did so on the first attempt.

\section*{Acknowledgements}

This work is part of the DIGPHAT project which was supported by a grant from the French government, managed by the National Research Agency (ANR), under the France 2030 program, with reference ANR-22-PESN-0017. KG would like to acknowledge the support of the National Science Centre, Poland, under grant number 2021/43/O/ST6/02805.

\section*{Data and Code availability}

All code used to generate the results in this study, along with configuration files, evaluation scripts, and preprocessed data sets, is publicly available in a dedicated GitHub repository: \url{github.com/ChristopheMuller/benchmark}. The repository also includes detailed instructions for reproducing the simulations, as well as tools for visualizing the results.

\clearpage

\newpage

\renewcommand{\thefigure}{\thesection.\arabic{figure}}  
\setcounter{figure}{0}  

\renewcommand{\thetable}{\thesection.\arabic{table}}  
\setcounter{table}{0}  

\begin{appendices}

\clearpage
\section{Methods}

\subsection{Acquisition of imputation methods}

We collected all imputation algorithms that we could find across packages, public repositories and scripts in Python and R, targeting general use cases. We focused on methods with publicly available implementations that could be applied in a wide range of scenarios without requiring domain-specific adjustments. Thus, we excluded methods that met any of the following criteria:
\begin{itemize}
    \item no implementation was publicly available;
    \item the method required at least one completely observed column;
    \item additional domain-specific metadata was necessary to perform the imputation;
    \item the method failed systematically across all scenarios (produced errors in 100\% of cases);
    \item the method was designed for MNAR (Missing Not At Random) scenarios, for example for censored data imputation;
    \item the method modified the original data in undesirable ways, such as removing outliers or applying preprocessing that conflicted with the goal of imputing missing values.
\end{itemize}

\subsection{Acquisition of data sets}

We selected a collection of data sets for imputation based on the following criteria:
\begin{itemize}
    \item Public availability: All data sets used in our study are publicly accessible.
    \item Sufficient size and variation: We excluded data sets that were too small to yield stable estimates. Our selection includes a variety of settings with differing sample sizes and numbers of variables, to reflect the diversity encountered in practical applications.
    \item Use in previous works: Preference was given to data sets that have been used in prior benchmarking studies or published analyses. This helped ensure that the data were of reasonable quality, including factors such as appropriate variable scaling, lack of strong multicollinearity, and a manageable number of missing values.
    \item General-purpose data sets: We focused on data sets that are not tied to a specific application area or task. 
\end{itemize}



\subsection{Pre-processing of data sets}

We did not modify the original data sets beyond minimal preprocessing. Specifically, we only removed constant or duplicate variables and, when necessary, adjusted column names. In addition, character variables were converted to factors, and numeric factors were recoded as appropriate to meet the requirements of the imputation methods (See \Cref{fig:categorical_methods}). Here, due to an incorrect hard-coded conversion of factor variables into numeric values using the \texttt{as.numeric} function by some of the \textsf{R} methods, we opted to reassign the levels of all factor variables to consecutive natural numbers $1, 2, 3, \ldots$ prior to imputation. This approach eliminated the risk of having different categories before and after imputation.

\subsection{Evaluation metrics}\label{Evaluationmetricsdetails}

Here we give further details on the evaluation metrics used for artificial and real missingness. Let in the following for $x \in \R^d$ $\| x \|_{2}$ be the Euclidean norm.

\subsubsection{Artificial missingness}

In the controlled experiment, we assess the performance of imputation methods by measuring the distributional distance between the imputed data set and the original complete data. To this end, we use the energy distance \cite{EnergyDistance}, defined as follows:
$$e^2(X, Y) = 2\mathbb{E} \|X - Y\|_{2} - \mathbb{E}\|X - X'\|_{2} - \mathbb{E}\|Y - Y'\|_{2},$$
where $X$ and $Y$ are random variables (here observed and imputed) and $d$ denotes length of a vector. Thus, if $X_i$, $i=1,\ldots,n$ are the observations of the original (fully observed) data set and $Y_i$, $i=1,\ldots,n$ are the imputations, then we calculate
\[
e_n^2(\mathbf{X}, \mathbf{Y}) = \dfrac{1}{n^2}\left( 2\sum_{i = 1}^n \sum_{j = 1}^n \|X_i - Y_j\|_{2} - \sum_{i = 1}^n \sum_{j = 1}^n \|X_i - X_j\|_{2} - \sum_{i = 1}^n \sum_{j = 1}^n \|Y_i - Y_j\|_{2}\right)
\]
using the \texttt{energy} \textsf{R} package \citep{energy}. 

Before computing the energy distance, we standardized all variables in both imputed and fully observed data sets using the column-wise means and standard deviations estimated from the fully observed data set \emph{prior to imposing missingness}. This ensures that the distance measure is not dominated by variables with large natural variance and is comparable across data sets. That is, for each variables $X_j$ and $Y_j$, we obtained scaled variables $\tilde{X}_j$ and $\tilde{Y}_j$ as follows:
$$\tilde{X}_j = \frac{X_j - \mu_j}{\sigma_j}  \quad \text{and  } \quad\tilde{Y}_j = \frac{Y_j - \mu_j}{\sigma_j}$$
where
$$
\mu_{j} = \frac{1}{n} \sum_{i=1}^n X_{ij},\quad \sigma_{j}^2 = \frac{1}{n-1} \sum_{i=1}^n (X_{ij} - \mu_j)^2$$
are mean and variance calculated based on the complete data set. This is what we refer to as \emph{standardized energy distance}. In order to be able to refer to other benchmarks, we also calculated pointwise evaluation metrics such as NRMSE (Normalized Root Mean Squared Error) and MPE (Mean Percentage Error)
$$NRMSE = \sqrt{\frac{1}{k}\left(\sum_{i=1}^{k}(\tilde{X}_i - \tilde{Y}_i)^2\right)}, \quad MPE = \frac{1}{k} \sum_{i=1}^{k}\left(\frac{X_i - Y_i}{X_i}\right) \cdot 100$$
where $k$ is the number of missing values and $\tilde{X}$ and $\tilde{Y}$ are standardized.

\subsubsection{Real missingness}

\begin{figure}
    \centering
    \includegraphics[width=1\linewidth]{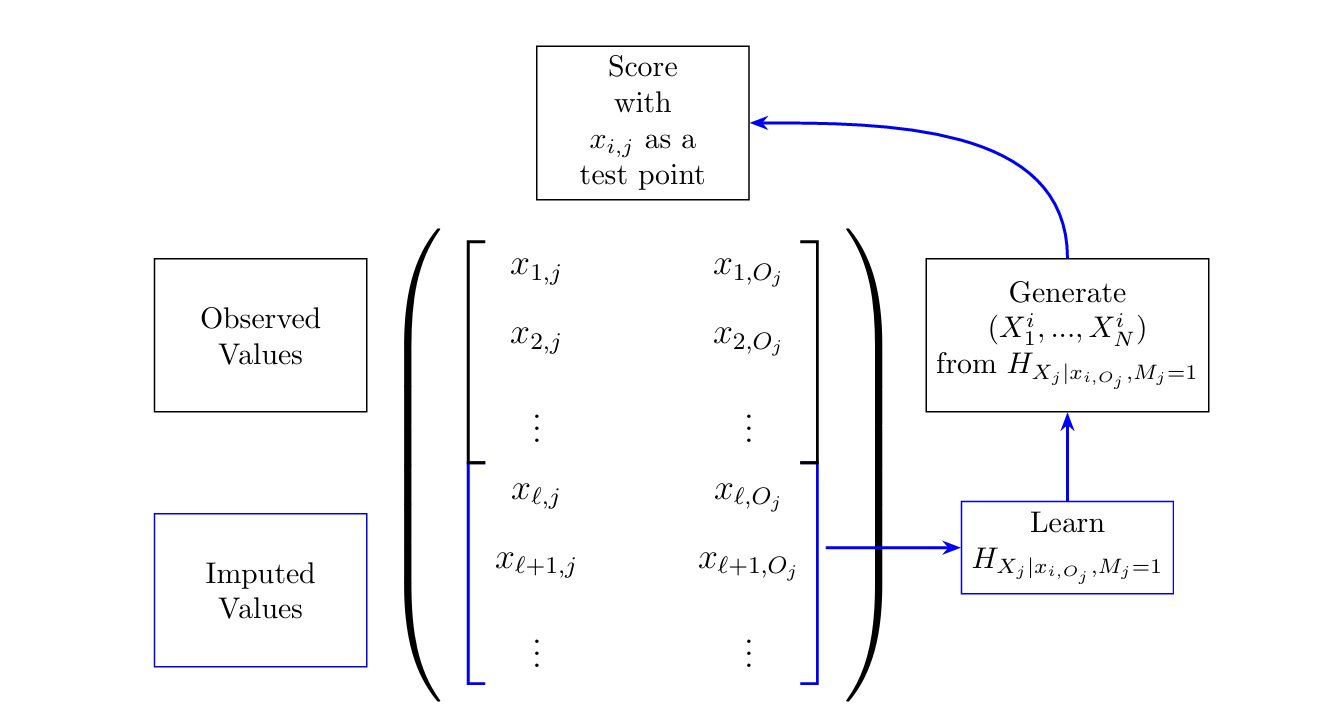}
    \caption{Illustration of the energy-I-Score, taken from \cite{näf2025Iscore}.}
    \label{fig:Scoreillustration}
\end{figure}

To evaluate imputation quality under real missingness, we use the \textit{energy-I-Score} implemented in the \texttt{miceDRF} package (\url{https://github.com/KrystynaGrzesiak/miceDRF}). 
It utilizes the energy score (see e.g., \cite{gneiting}) by comparing the imputation distribution $H$ with the real distribution as follows: For each variable $j$ in set set $\mathcal{S} \subset \{1,\ldots, d\}$, we consider the biggest set of variables $X_{O_j}$ such that the variables $X_{O_j}$ are observed, whenever $X_j$ is observed. Let in the following the variable $M_j$ indicate whether $X_j$ is observed ($M_j=0$) or not ($M_j=1$). The goal is to compare the conditional imputation distribution $H_{X_j \mid X_{O_j}, M_j = 1}$ to the observed conditional distribution $P_{X_j \mid X_{O_j}, M_j = 0}$. Formally, for each variable $j \in \mathcal{S}$ with missing values, the expected score is defined as
\begin{align*}
  &S^j_{\textrm{\tiny NA}}( H, P ) = \nonumber\\
  &\E_{X_{O_j} \sim P_{X_{O_j}\mid M_j=0}}\Big[   \E_{\substack{X \sim H_{X_j \mid X_{O_j}, M_j=1}\\Y \sim P_{X_j \mid X_{O_j}, M_j=0}}}[ | X - Y |] - \frac{1}{2} \E_{\substack{ X \sim H_{X_j \mid X_{O_j}, M_j=1}\\ X' \sim H_{X_j \mid X_{O_j}, M_j=1}}}[| X-X' |] \Big].
\end{align*}
The overall energy-I-Score is the average across all variables with missing values:
\[
S^{NA}(H, P) = 
\frac{1}{|\mathcal{S}|}
\sum_{j \in \mathcal{S}} 
S^{NA}_j(H, P).
\]
In practice, the expectations are approximated by repeated imputations 
$\tilde{X}^{(1)}, \ldots, \tilde{X}^{(N)}$ drawn from $H_{X_j \mid X_{O_j}, M_j = 1}$, leading to the following estimator:
\[
\hat{S}^{NA}_j(H, P) = 
\frac{1}{|L_j|} 
\sum_{i \in L_j}
\left[
\frac{1}{2N^2} 
\sum_{l,\ell=1}^{N} 
\left| 
\tilde{X}_{ij}^{(l)} - \tilde{X}_{ij}^{(\ell)} 
\right|
- 
\frac{1}{N} 
\sum_{l=1}^{N} 
\left| 
\tilde{X}_{ij}^{(l)} - x_{ij} 
\right|
\right].
\]
We also apply the energy-I-Score to categorical variables. In this case, $X_j$ is one-hot encoded into binary variables $X_{j_1}, \ldots, X_{j_p}$, which are treated jointly as a multivariate vector. 
The score is then computed using the same formula as for continuous variables, replacing scalar values with their one-hot representations:
\[
\widehat{S}^j_{\textrm{\tiny NA}}( H, P ) =
\frac{1}{|L_j|} 
\sum_{i \in L_j} 
\left(
\frac{1}{2N^2} 
\sum_{l,\ell=1}^N 
\| \tilde{\mathbf{X}}_{i,j}^{(l)} - \tilde{\mathbf{X}}_{i,j}^{(\ell)} \|_2
- 
\frac{1}{N} 
\sum_{l=1}^N 
\| \tilde{\mathbf{X}}_{i,j}^{(l)} - \mathbf{x}_{i,j} \|_2
\right).
\]
Figure \ref{fig:Scoreillustration} illustrates the scoring for a fixed $j$. For further details we refer to \cite{näf2025Iscore}.

\subsection{Implementation details}

Most of the implementation was done in \textsf{R}, where the entire simulation pipeline was supervised and parallelized using the \texttt{targets} package \citep{targets}.  For running imputation methods implemented in \textsf{Python}, we used the \texttt{reticulate} package \citep{reticulate}, which allowed seamless integration between both languages within the same pipeline. All package versions used in our analysis can be found in the \texttt{renv.lock} configuration file generated using the \texttt{renv} package \citep{renv}, ensuring full reproducibility of the computational environment.




\subsection{Imputation}

\subsubsection{Scaling}

We did not apply feature scaling prior to imputation. While scaling is often a recommended preprocessing step for many machine learning algorithms, it becomes non-trivial in the presence of missing values. Standard scaling approaches require complete observations to compute statistics such as means or standard deviations. Applying them before imputation would require either dropping incomplete cases (which contradicts our setting) or relying on partial data, introducing bias or inconsistency. Furthermore, some imputation methods implicitly handle scaling through their internal modeling (e.g., PCA or tree-based methods), making external scaling unnecessary or even detrimental.

\subsubsection{Numerical accuracy}

We systematically evaluated the output of each imputation method for correctness. In cases where methods returned results that were nearly identical to the original data, we allowed for minimal numerical differences due to floating-point precision. Specifically, we tolerated discrepancies smaller than $1.5 \times 10^{-5}$, which is slightly higher than commonly adopted values in numerical validation. This more generous threshold was chosen to avoid penalizing methods for negligible numerical differences—particularly those arising from conversions between \textsf{R} and \textsf{python}. If all imputed values differed from the original ones by less than this margin, we considered the method as not having meaningfully modified the data, and the run was flagged accordingly.

\subsubsection{Imputation of categorical variables}
\label{sec:cat_imputation}

\begin{figure}[!h]
    \centering
    \includegraphics[width=1 \linewidth]{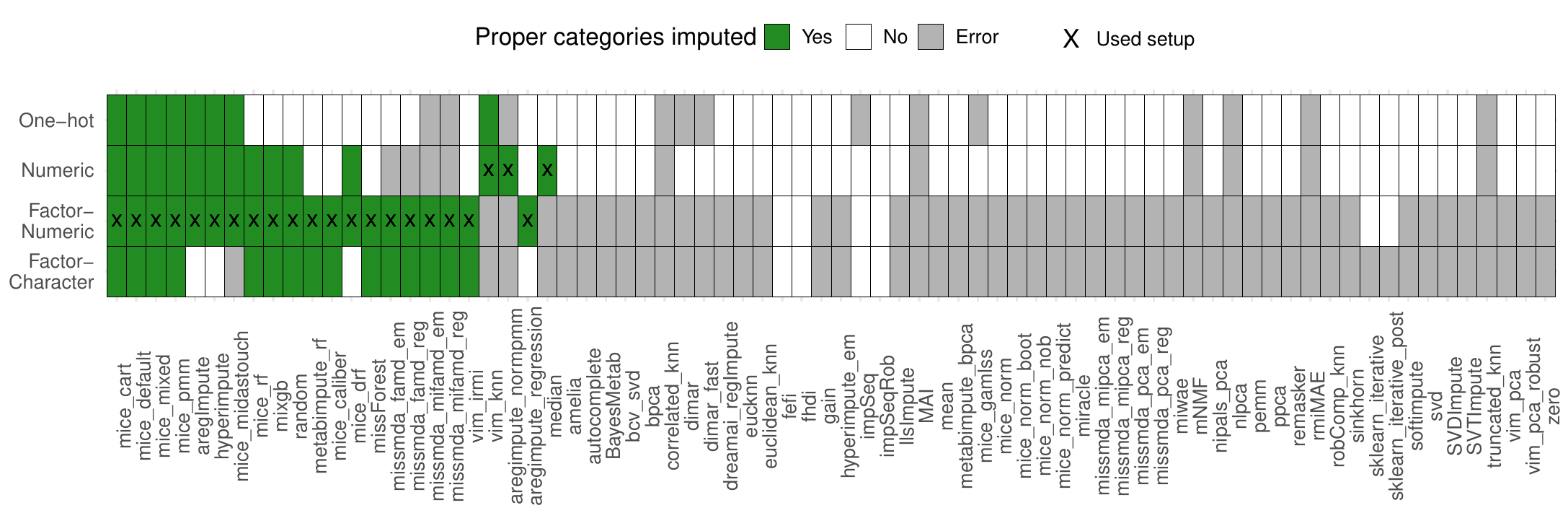}
    \caption{\textit{Selected imputation setup for mixed-type data. Green color indicates that the method successfully imputes categorical variables in the given format. White color means that the method completed imputation, but the resulting categories did not match the expected ones. Grey color indicates that the method failed with an error. The \textbf{X} symbol marks the format selected for the benchmark.}}
    \label{fig:categorical_methods}
\end{figure}

To check whether imputation methods handle categorical variables correctly, we designed a small set of diagnostic cases representing common encoding formats: integer, factor (with numeric or character levels), and one-hot encoded variables. For each case, we applied all the available imputation methods and evaluated whether the output respected the expected structure. Specifically, we checked:

\begin{itemize}
    \item for numeric or factor formats, whether imputed categorical variables retained valid levels (from $\G$),
    \item for one-hot encodings, whether each observation had exactly one active category and whether dummy variables only contained binary values.
\end{itemize}
Next, based on performance, we marked a method as capable of imputing mixed-type data if it passed all checks for at least one encoding format. Some methods failed with an error. The results are presented in a  \Cref{fig:categorical_methods}. In total, \nrofmethodscat methods successfully imputed categorical variables. Among them, three methods were evaluated in the benchmark using a numeric encoding of categorical variables, while the remaining methods were applied with a factor or integer-coded format.

Let us note that among all the methods included, two were specifically designed for mixed-type data: \texttt{missmda\_mifamd\_em} and \texttt{missmda\_mifamd\_reg}. These methods are not applicable to purely numerical data sets and consistently failed in such settings, which aligns with their intended design. We thus chose to exclude them from the numerical-only scenarios.

\clearpage
\section{Results}

\subsection{Detailed results}

\begin{figure}[h]
    \centering
    \includegraphics[width=1 \linewidth]{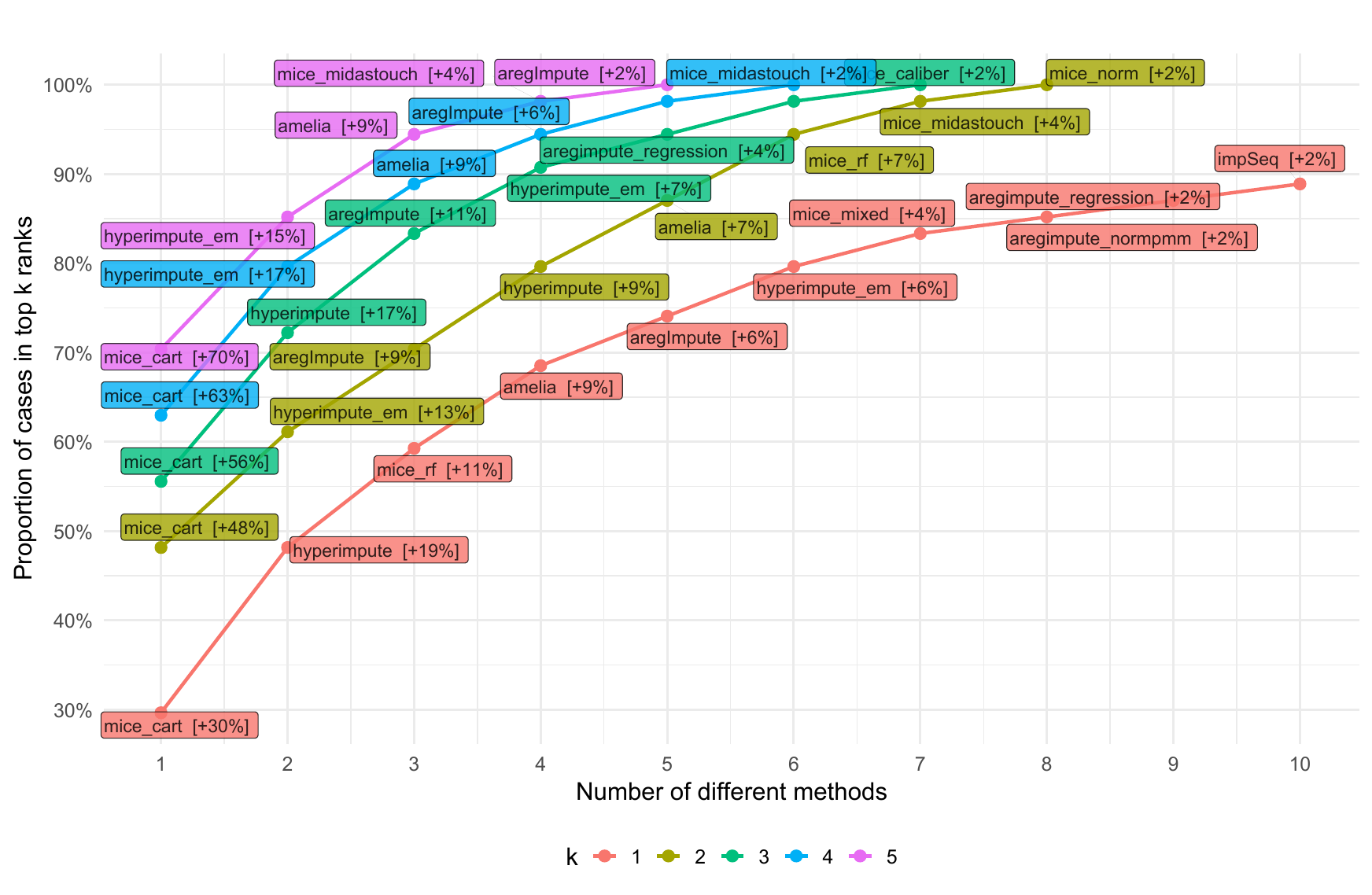}
    \caption{Proportion of imputation scenarios for numerical data with artificial missingness where at least one method in the selected group achieves a rank within the top-k, plotted as a function of the number of included methods. Curves correspond to different values of k (from 1 to 5).}
    \label{fig:top-k}
\end{figure}

\begin{figure}[h]
    \centering
    \includegraphics[width=1 \linewidth]{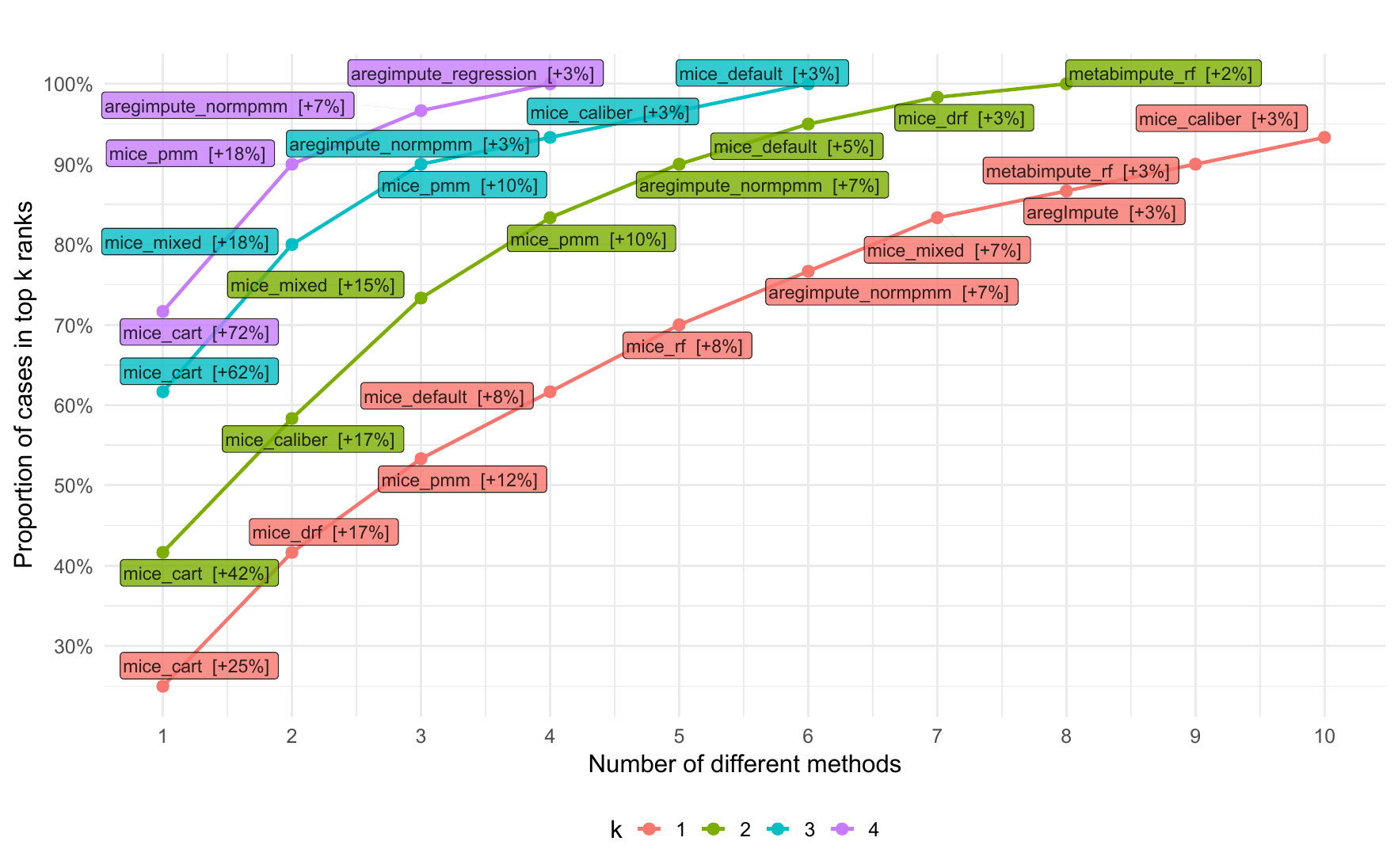}
    \caption{Proportion of imputation scenarios for mixed data with artificial missingness where at least one method in the selected group achieves a rank within the top-k, plotted as a function of the number of included methods. Curves correspond to different values of k (from 1 to 4).}
    \label{fig:top-k_cat}
\end{figure}

\begin{figure}[h]
    \centering
    \includegraphics[width=1 \linewidth]{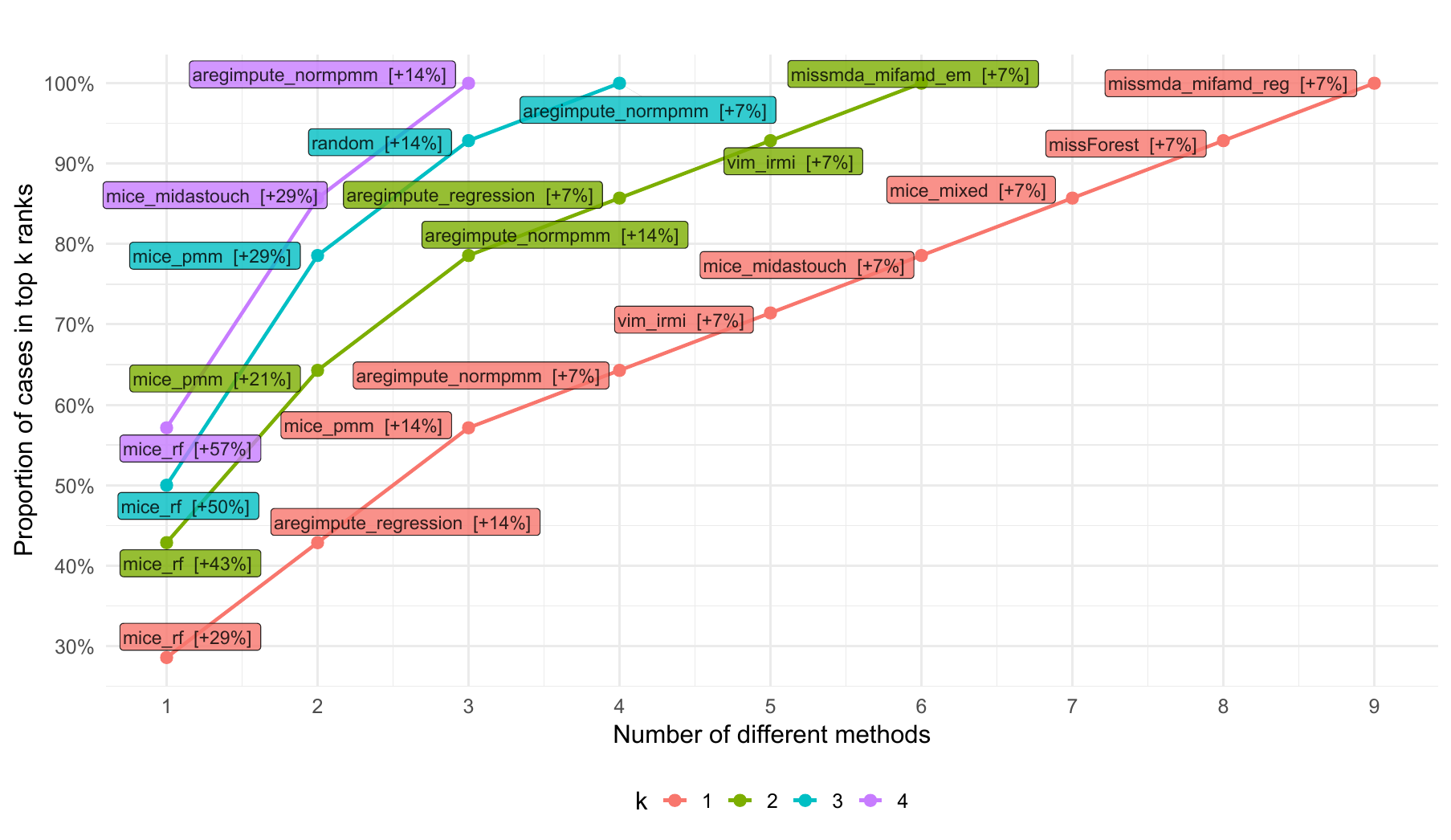}
    \caption{Proportion of imputation scenarios for numerical and mixed data with real missingness where at least one method in the selected group achieves a rank within the top-k, plotted as a function of the number of included methods. Curves correspond to different values of k (from 1 to 4).}
    \label{fig:top-k_inc}
\end{figure}

\begin{figure}[p]
    \centering
    \includegraphics[angle=270, width=\linewidth]{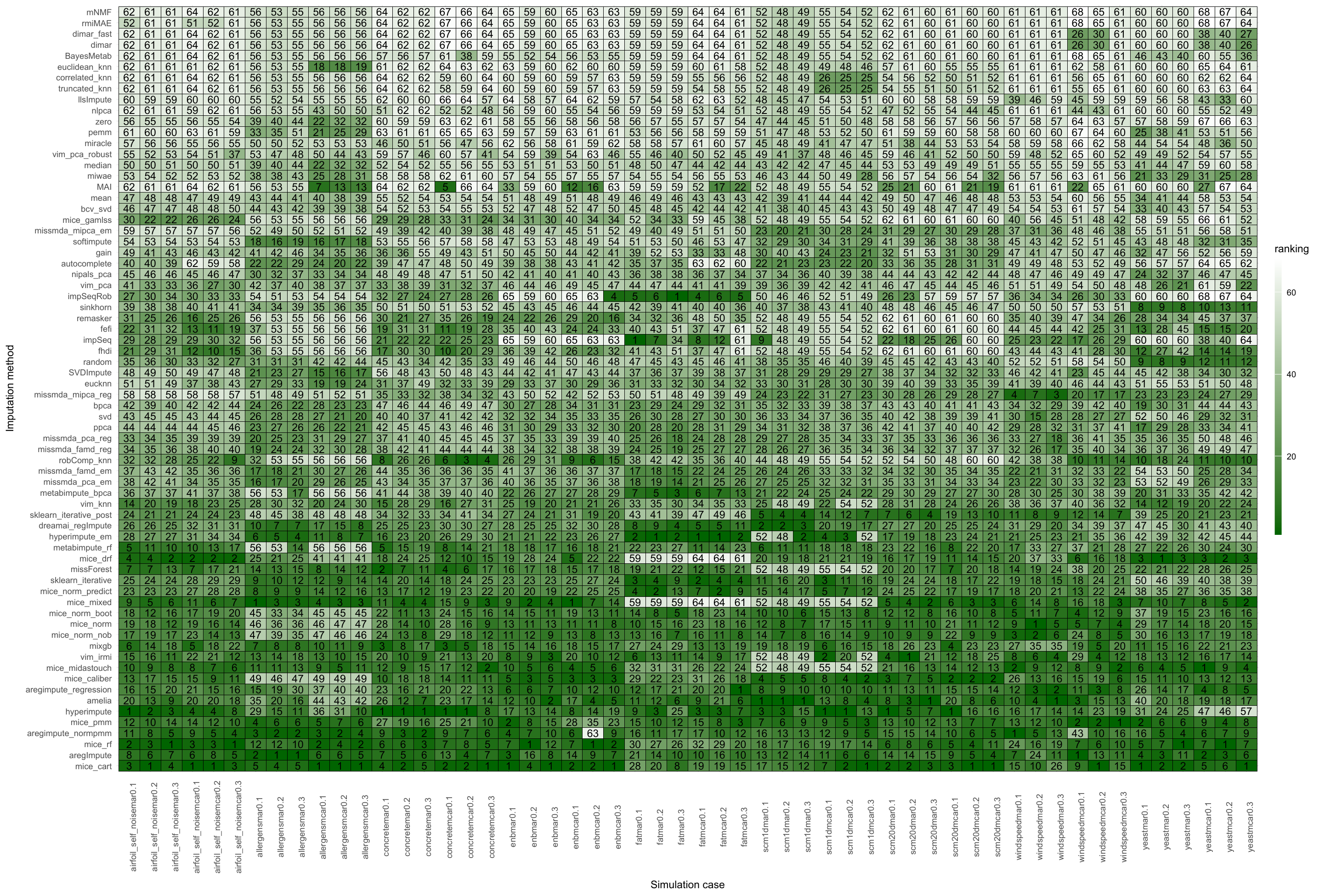}
    \caption{Detailed rankings on numerical data sets with artificial missingness.}
    \label{fig:RealMissing_Shrek}
\end{figure}

\begin{figure}[p]
    \centering
    \includegraphics[width=\linewidth]{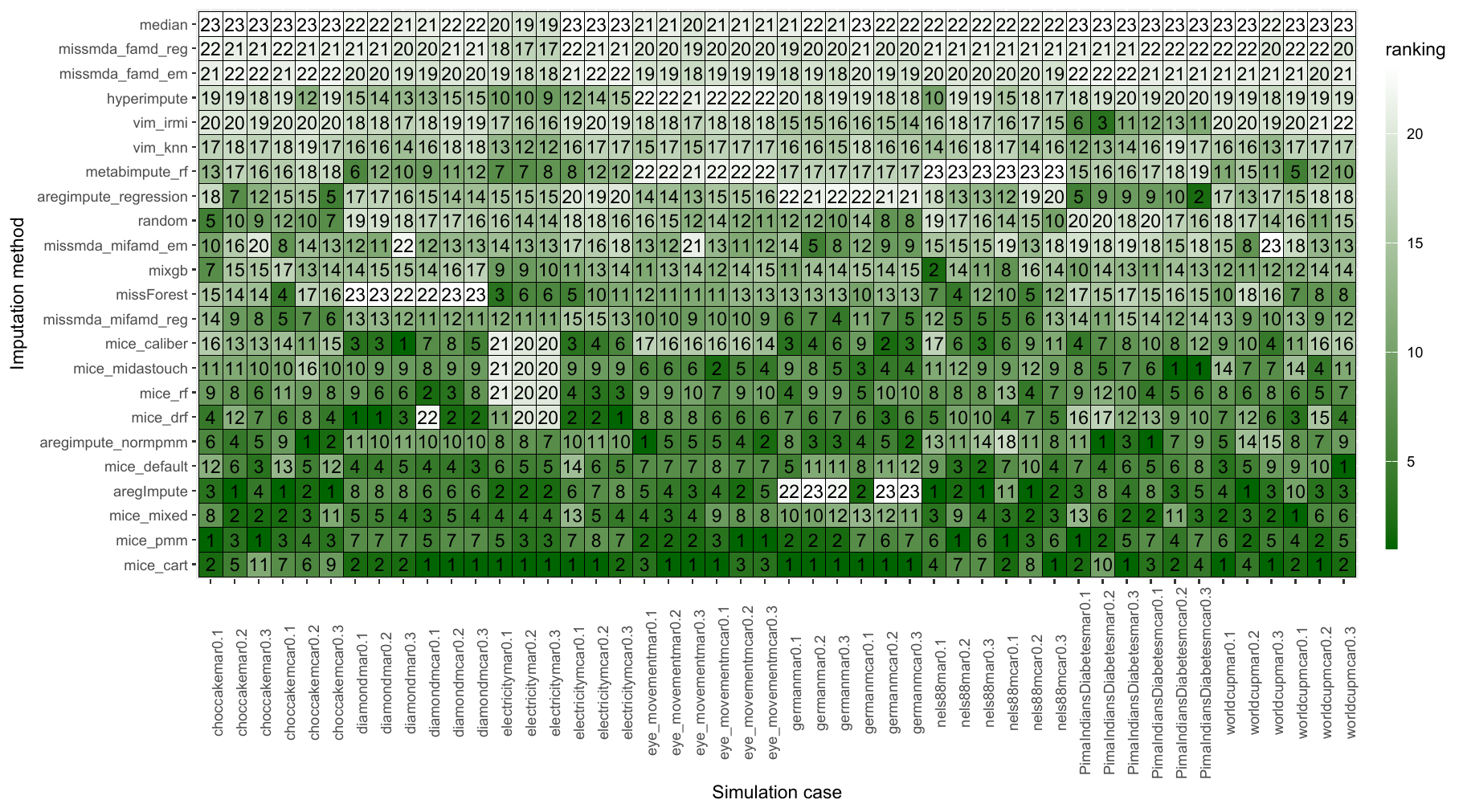}
    \caption{Detailed rankings on mixed data sets with artificial missingness.}
    \label{fig:shrek_cat}
    \label{fig:RealMissing_Shrek_cat}
\end{figure}

\begin{figure}[p]
    \centering
    \includegraphics[width=\linewidth]{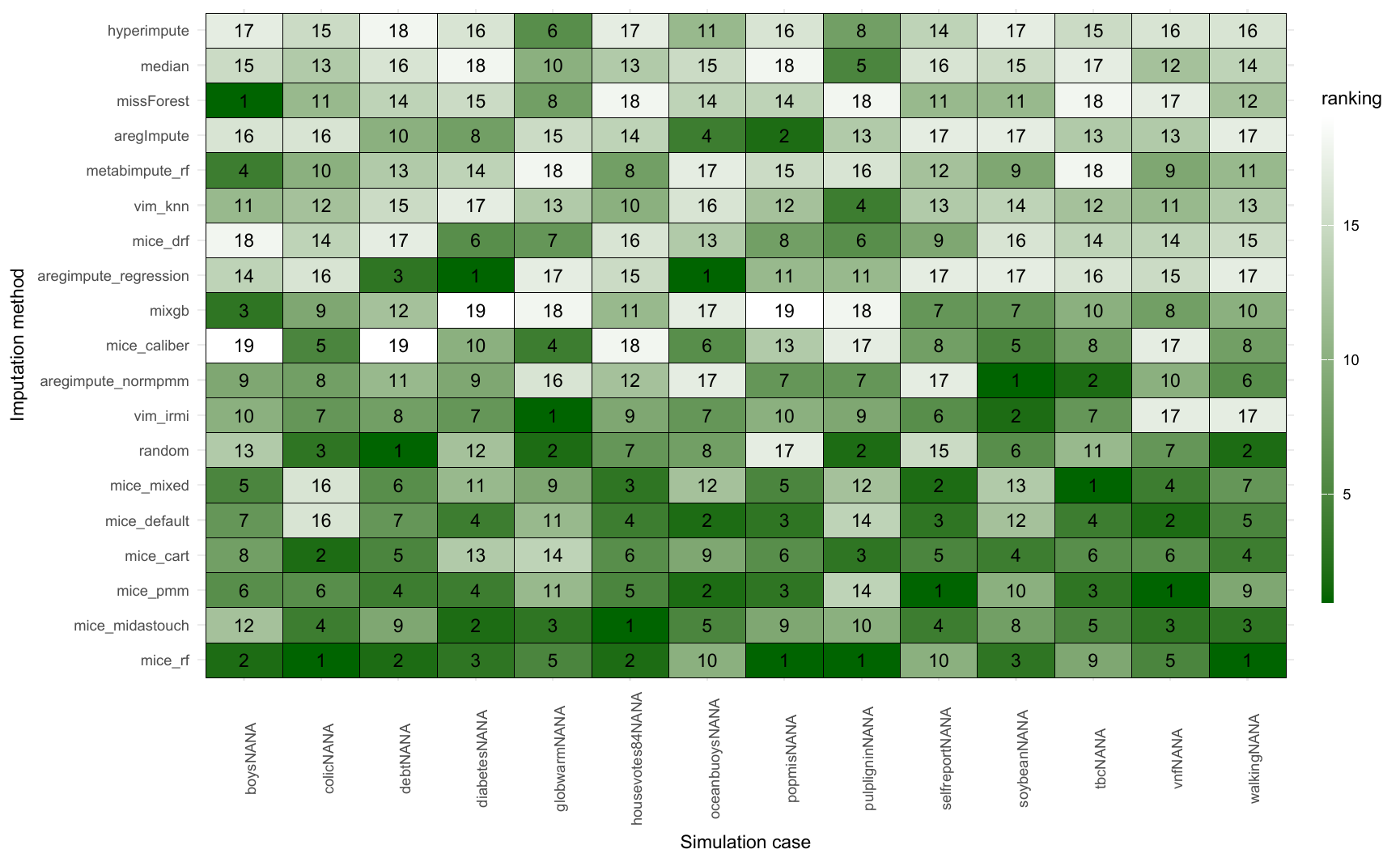}
    \caption{Detailed rankings on numeric and mixed data sets with real missingness.}
    \label{fig:shrek_inc}
    \label{fig:RealMissing_Shrek_inc}
\end{figure}

\begin{figure}
    \centering
    \includegraphics[width=01 \linewidth]{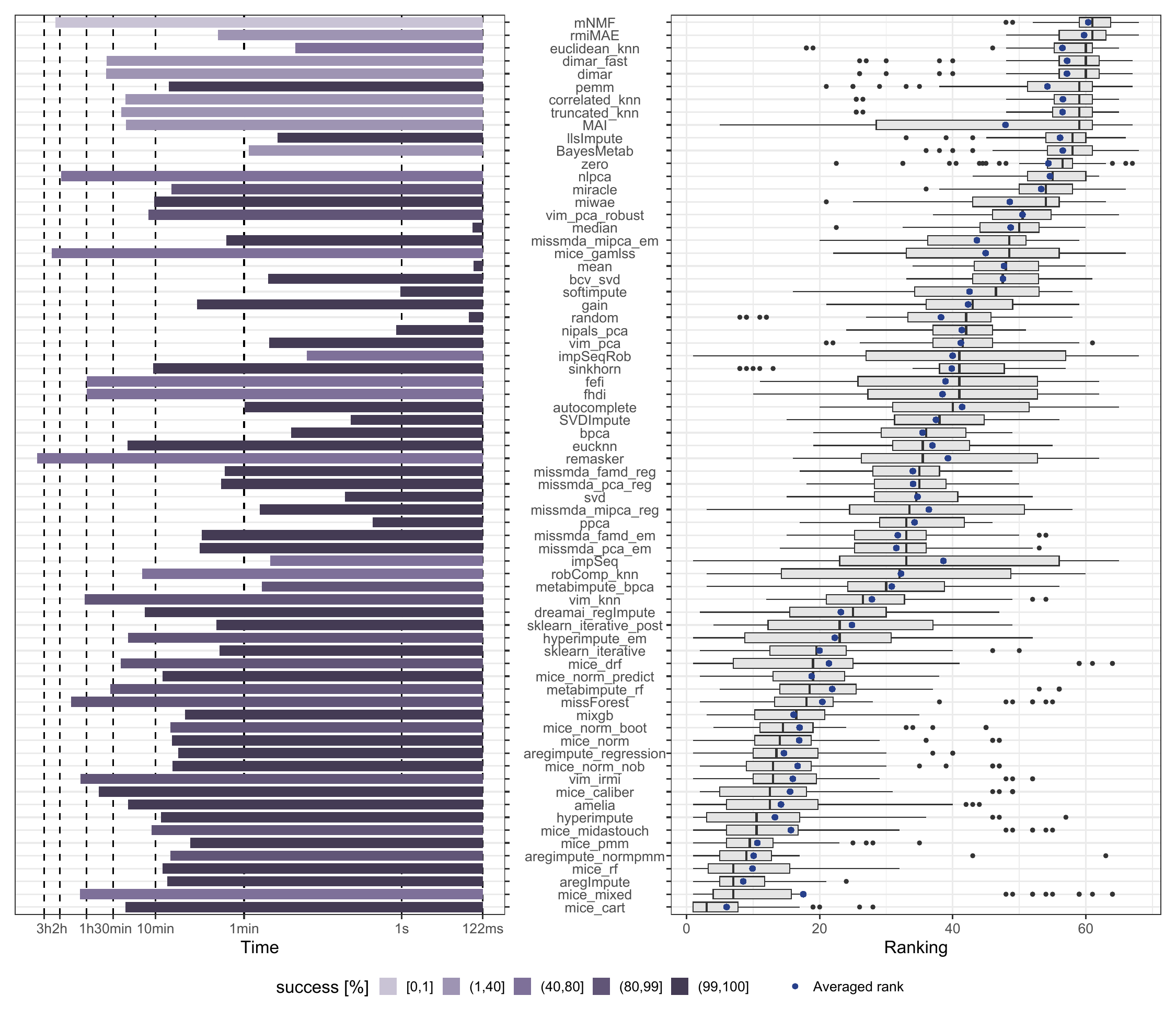}
    \caption{Results for artificial missingness, on numerical data sets. Left: averaged evaluation time. Right: position in ranking averaged over 2 replications of amputation. The ordering corresponds to median rank.}
    \label{fig:FakeMissing_numericalData_median}
\end{figure}

In this section we provide additional and more detailed results. First, Figures \ref{fig:top-k}, \ref{fig:top-k_cat} show the top performers for numerical and mixed data respectively. Figure \ref{fig:top-k_inc} contains similar results for the simulation cases with real missingness. Moreover, the Figures \ref{fig:RealMissing_Shrek}, \ref{fig:RealMissing_Shrek_cat} and \ref{fig:RealMissing_Shrek_inc}  contain the detailed rankings that were summarized in the manuscript. The Figure \ref{fig:FakeMissing_numericalData_median} corresponds to \ref{fig:FakeMissing_numericalData} but instead of mean, the methods are ordered using median.

\subsection{Stability}\label{Sec_Stability}

\begin{figure}[h]
    \centering
    \includegraphics[width=1 \linewidth]{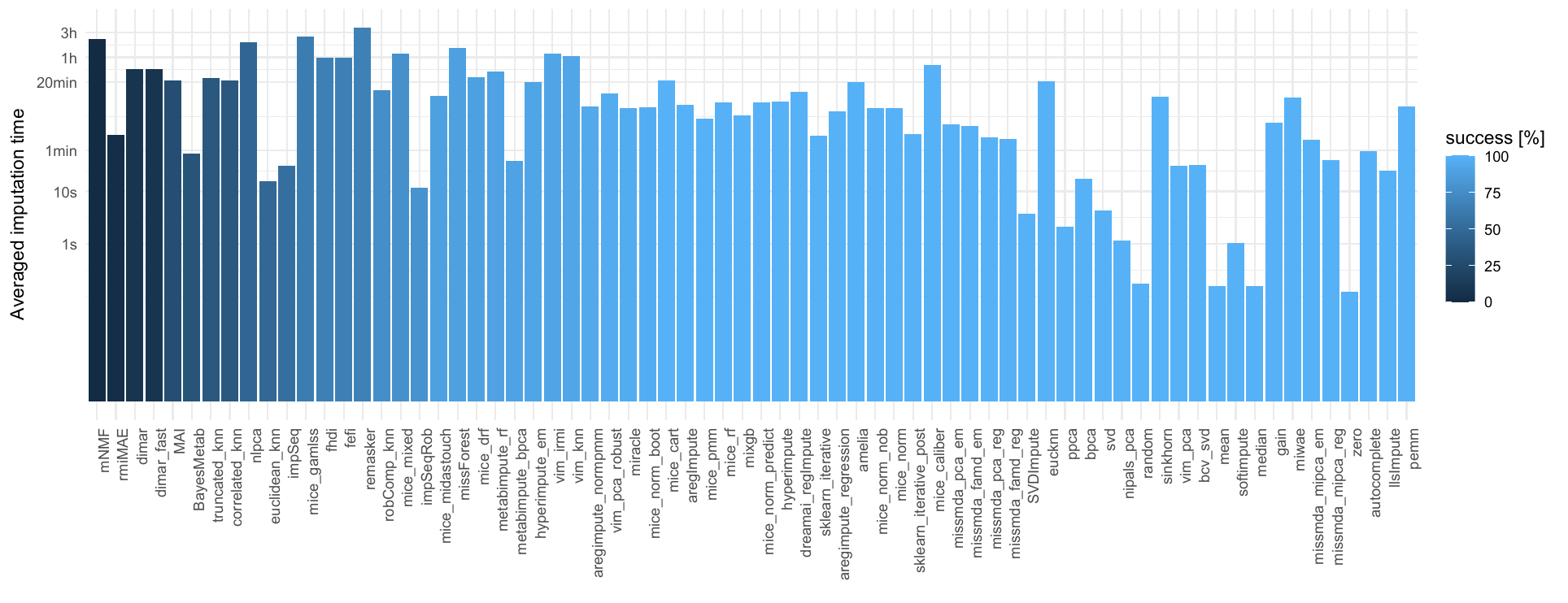}
    \caption{Averaged imputation time on numerical data sets with artificial missingness per method (log scale), ordered by overall success rate.}
    \label{fig:times_success}
\end{figure}

\begin{figure}[h]
    \centering
    \includegraphics[width=1 \linewidth]{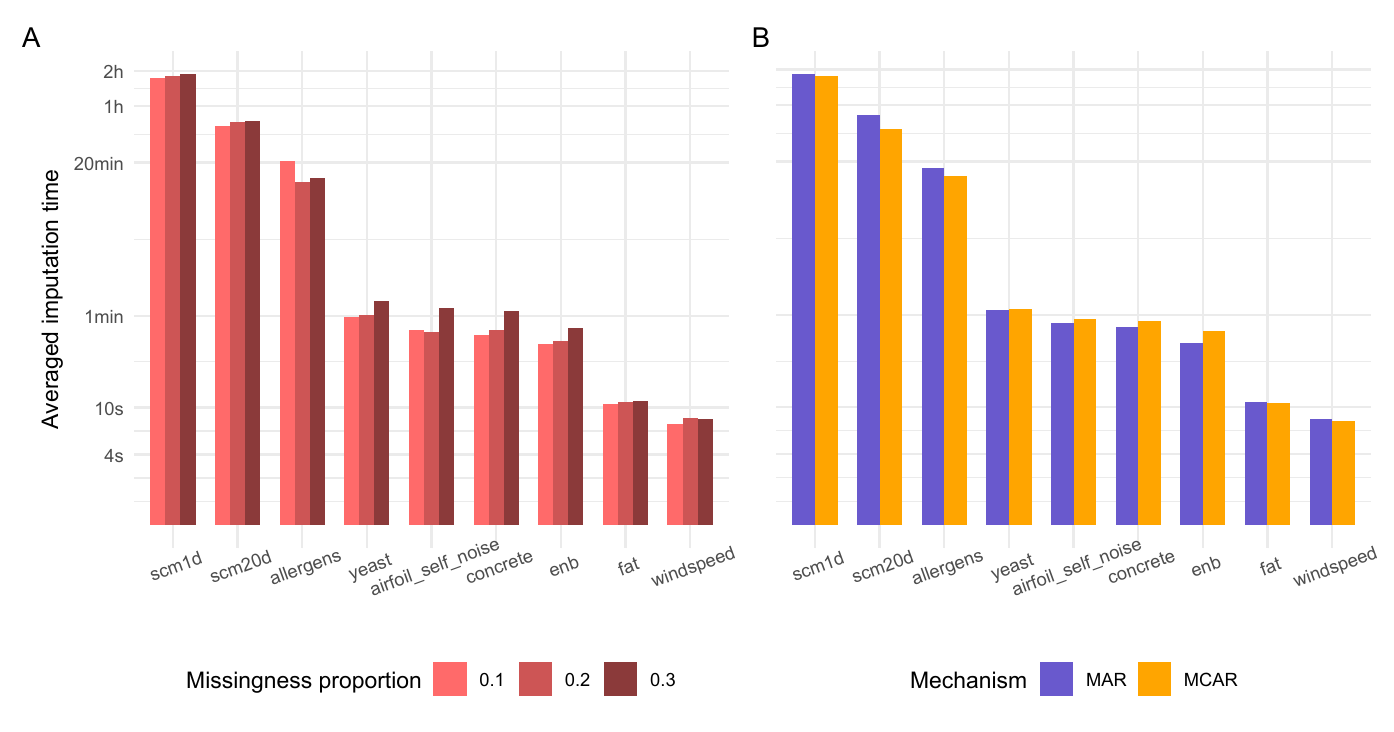}
    \caption{Averaged imputation time (log10 scale) across on numerical data sets with artificial missingness depending on missingness characteristics. (A) Imputation time stratified by the proportion of missing values (10\%, 20\%, 30\%). (B) Imputation time stratified by the missingness mechanism (MCAR, MAR).}
    \label{fig:times_cases}
\end{figure}

\begin{figure}[h]
    \centering
    \includegraphics[width=1 \linewidth]{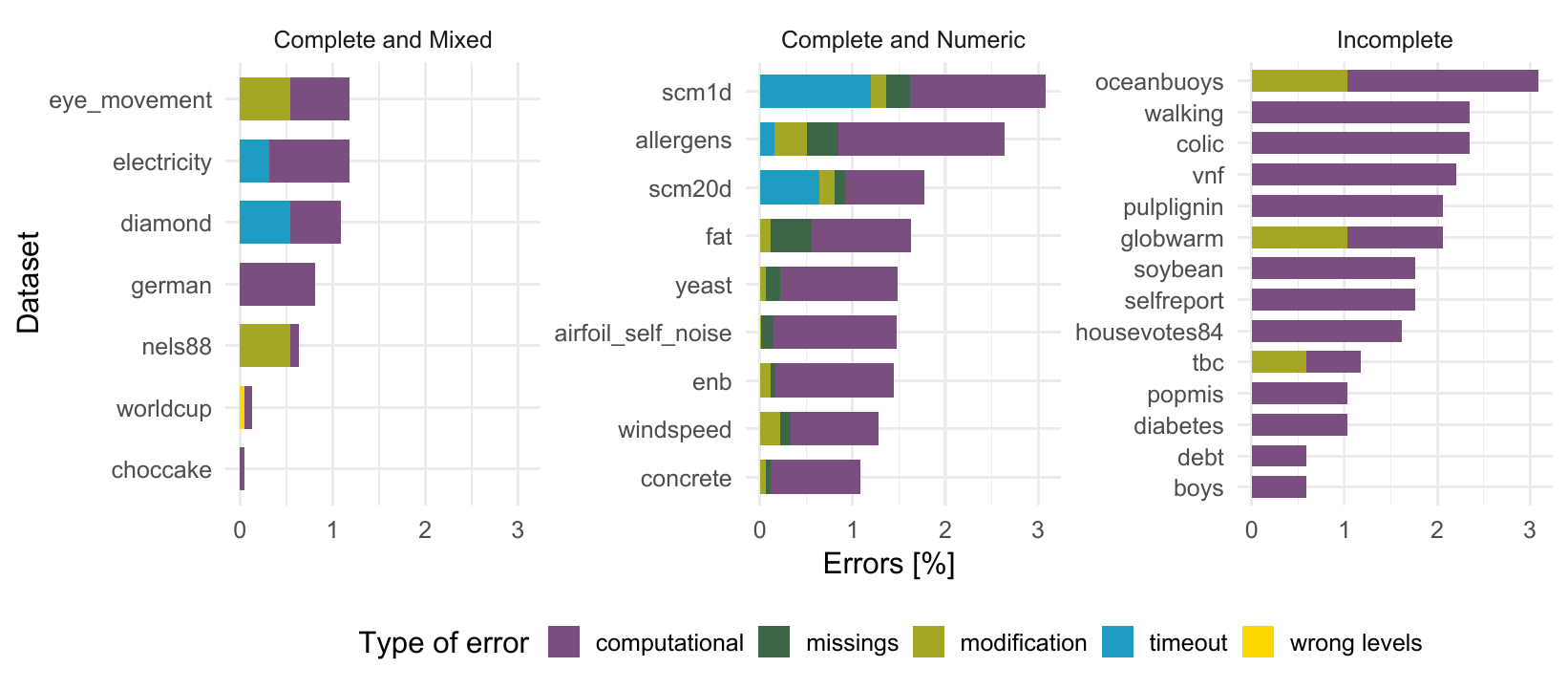}
    \caption{Errors produced by all methods, by data set and simulation scenario.}
    \label{fig:errors_datasets}
\end{figure}

Here we detail the numerical stability of the methods, as well as their imputation times in Figures \ref{fig:times_success}, \ref{fig:times_cases} and  \ref{fig:errors_datasets}.

\subsection{Comparison of measures}

In this section we go beyond the standardized energy distance and show the rankings according to several different measures, including the energy distance.

\begin{figure}[p]
    \centering
    \includegraphics[width=1\linewidth]{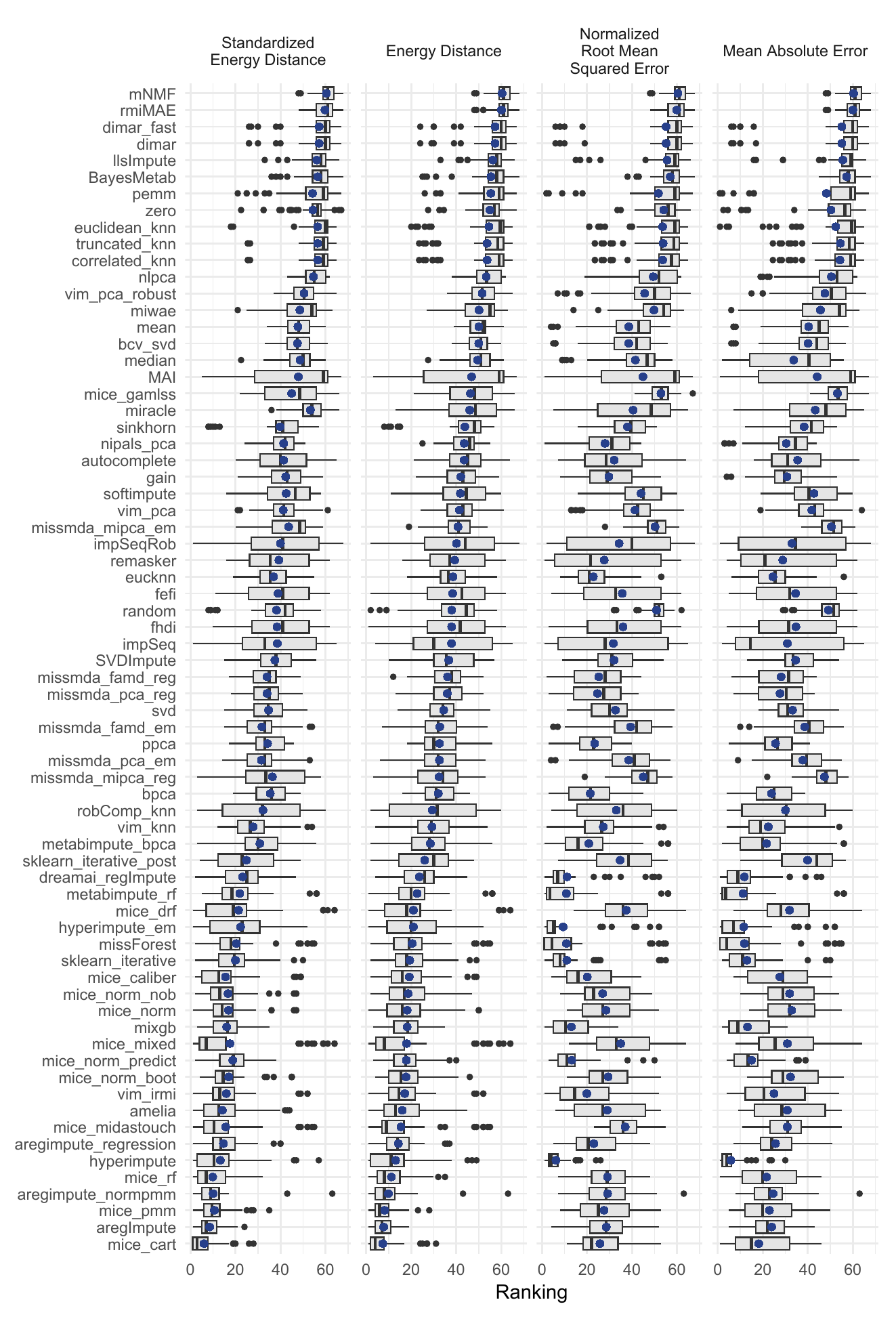}
    \caption{A comparison of four evaluated imputation quality metrics. The methods are ordered according to energy distance.}
    \label{fig:measures}
\end{figure}

\end{appendices}

\clearpage

\bibliographystyle{apalike}
\bibliography{biblioMAR}

\end{document}